\documentclass[usenatbib,fleqn]{mnras}
\usepackage{graphicx,hyperref,amsmath,amssymb,xspace}
\usepackage{newtxtext,newtxmath}  % more compact font

\usepackage[dvipsnames]{xcolor}
\usepackage{tikz}
\usetikzlibrary{svg.path}
\definecolor{orcidlogocol}{HTML}{A6CE39}
\tikzset{orcidlogo/.pic={
 \fill[orcidlogocol] svg{M256,128c0,70.7-57.3,128-128,128C57.3,256,0,198.7,0,128C0,57.3,57.3,0,128,0C198.7,0,256,57.3,256,128z};
 \fill[white] svg{M86.3,186.2H70.9V79.1h15.4v48.4V186.2z}
 svg{M108.9,79.1h41.6c39.6,0,57,28.3,57,53.6c0,27.5-21.5,53.6-56.8,53.6h-41.8V79.1z M124.3,172.4h24.5c34.9,0,42.9-26.5,42.9-39.7c0-21.5-13.7-39.7-43.7-39.7h-23.7V172.4z}
 svg{M88.7,56.8c0,5.5-4.5,10.1-10.1,10.1c-5.6,0-10.1-4.6-10.1-10.1c0-5.6,4.5-10.1,10.1-10.1C84.2,46.7,88.7,51.3,88.7,56.8z};
}}
\newcommand\orcidicon[1]{\href{https://orcid.org/#1}{
\begin{tikzpicture}[yscale=-0.04,xscale=0.04,transform shape]
\pic{orcidlogo};
\end{tikzpicture}
}}

\title{Dear Magellanic Clouds, welcome back!}

\author[E.Vasiliev]{
Eugene Vasiliev~\orcidicon{0000-0002-5038-9267}$^{1}$\thanks{E-mail: eugvas@protonmail.com}\\
$^1$Institute of Astronomy, Madingley road, Cambridge, CB3 0HA, UK}

\newcommand{\Gaia}{\textit{Gaia}\xspace}
\newcommand{\kms}{km\:s$^{-1}$\xspace}
\newcommand{\masyr}{mas\:yr$^{-1}$\xspace}
\renewcommand{\L}{$\mathcal L$}
\newcommand{\M}{$\mathcal M$}
\defcitealias{Vasiliev2023}{V23}

\date{Accepted 2023 August 25. Received 2023 August 14; in original form 2023 June 8}

\begin{document}
\label{firstpage}
\pagerange{437--456}\volume{527}\pubyear{2024}
\setcounter{page}{437}
\maketitle

\begin{abstract}
We propose a scenario in which the Large Magellanic Cloud (LMC) is on its second passage around the Milky Way. Using a series of tailored $N$-body simulations, we demonstrate that such orbits are consistent with current observational constraints on the mass distribution and relative velocity of both galaxies. The previous pericentre passage of the LMC could have occurred 5--10~Gyr ago at a distance $\gtrsim 100$~kpc, large enough to retain its current population of satellites. The perturbations of the Milky Way halo induced by the LMC look nearly identical to the first-passage scenario, however, the distribution of LMC debris is considerably broader in the second-passage model. We examine the likelihood of current and past association with the Magellanic system for dwarf galaxies in the Local Group, and find that in addition to 10--11 current LMC satellites, it could have brought a further 4--6 galaxies that have been lost after the first pericentre passage. In particular, four of the classical dwarfs -- Carina, Draco, Fornax and Ursa Minor -- each have a $\sim50$\% probability of once belonging to the Magellanic system, thus providing a possible explanation for the ``plane of satellites'' conundrum.
\end{abstract}


\section{Introduction}

The Magellanic clouds are the most prominent and best-known satellites of our Galaxy, known since ancient times, but their importance for the Milky Way dynamics only began to be appreciated in the last decade. The LMC mass is only a few times smaller than that of the Milky Way, and it is currently just passed the pericentre of its orbit at $\sim$50~kpc, thus its dynamical influence on our Galaxy appears to be quite substantial (see \citealt{Vasiliev2023}, hereafter \citetalias{Vasiliev2023}, for a review of the many facets of this interaction). For a long time, it was believed that these galaxies are long-term satellites of the Milky Way, completing many orbital periods in its lifetime \citep[e.g.,][]{Tremaine1976}. However, after first measurements of the proper motion (PM) of the LMC were published by \citet{Kallivayalil2006a} using two-epoch \textit{Hubble} space telescope (HST) data, it became clear that its tangential velocity is very high ($\gtrsim 300$~\kms) and possibly even exceeds the escape velocity of the Milky Way at that distance. Thus the scenario in which the Magellanic Clouds are on their first passage around the Galaxy \citep{Besla2007} became predominant. 

Subsequent analysis of additional HST observations \citep{Kallivayalil2013}, as well as independent PM measurements from the \Gaia satellite \citep{Helmi2018,Luri2021}, led to a downward revision of the LMC tangential velocity by a few tens \kms (see Table~2 in  \citetalias{Vasiliev2023} for a compilation of results and Figure~2 in that paper for the illustration of differences in the past orbit). Depending on the Milky Way mass distribution, it is not impossible to imagine that the Magellanic Clouds have completed more than one orbit during the Hubble time, but most studies in recent years remain focused on the first-passage scenarios. Meanwhile, the Galactic potential and total mass also became better constrained in the \Gaia era (see \citealt{Wang2020} for a recent review). A fiducial value of $M_\mathrm{MW}\simeq 10^{12}\,M_\odot$ with an optimistic uncertainty level of 20--30\% emerges from several independent methods, placing it on a lower end of the historically considered range of values and reinforcing the arguments in favour of a first-passage Magellanic orbit. Nevertheless, as discussed in Section~3.2 of \citetalias{Vasiliev2023}, the current level of uncertainty in both the Galactic potential and the LMC parameters is still not sufficient to firmly exclude either first or second-passage scenario, primarily because the orbit is only marginally bound, and even a small variation in total energy leads to dramatic changes in the inferred orbital period. 

In this paper, we consider the implications of the second-passage scenario, in which the LMC had a previous encounter with the Milky Way between 5 and 10~Gyr ago at a distance $\gtrsim 100$~kpc. In section~\ref{sec:simulations} we introduce the suite of $N$-body simulations of the Milky Way--LMC system. Then in Section~\ref{sec:analysis} we discuss the properties of the LMC orbit, distribution of its tidal debris, perturbations induced in the Milky Way, and its future fate. Section~\ref{sec:satellites} is devoted to the analysis of possible association of Local group dwarf galaxies with the Magellanic system. Finally, in Section~\ref{sec:discussion} we summarize our findings and outline open questions for future studies.

%%%%%%%%
\section{Simulations}  \label{sec:simulations}

\subsection{Initial models}  \label{sec:initialmodels}

We model both galaxies as pure $N$-body systems, neglecting any effect of the gas. This is justified by the expectation that the orbit is determined by the total gravitational potential, which is dominated by dark matter, with a secondary contribution of stars, and an even smaller contribution of gas. \citet{TepperGarcia2019} find that the addition of a hot gas corona in a hydrodynamical simulation does not appreciably change the orbit of the LMC compared to the purely collisionless case (although, of course, it affects the properties of the Magellanic gas stream, which we cannot study in our setup).

Both galaxies are set up as equilibrium models initially, using various tools from the stellar-dynamical toolbox \textsc{Agama} \citep{Vasiliev2019}. This is trivial for the LMC, which is represented by a spherical density profile with the distribution function computed from the Eddington inversion formula. We do not explicitly include the stellar disc or halo of the LMC, because these are sub-dominant in the total mass distribution (\citealt{vanderMarel2002} estimated the stellar mass to be $3\times10^9\,M_\odot$) and located sufficiently deep in the potential well to avoid being stripped. The density profile of the LMC is designed to match the circular-velocity curve in the inner few kpc determined from stellar kinematics \citep{Vasiliev2018} and the enclosed mass constraint of $1.7\times10^{10}\,M_\odot$ at 8.7~kpc from the LMC centre \citep{vanderMarel2014}, so the inner region of our LMC model effectively represents both stars and dark matter. The density follows the Navarro--Frenk--White (NFW) profile with a smooth but rather sharp cutoff:
\begin{equation}  \label{eq:rho_halo}
\rho_\mathrm{halo} \propto r^{-1}\,\big(1+r/r_\mathrm{s}\big)^{-2}\,\exp\big[-(r/r_\mathrm{c})^4\big].
\end{equation}

The Milky Way models consist of a spherical dark matter halo with the same tapered NFW density profile, and two stellar components: a spherical bulge with density profile
\begin{equation}  \label{eq:rho_bulge}
\rho_\mathrm{bulge} \propto (1+r/0.2\,\mbox{kpc})^{-1.8}\; \exp\big[-(r/1.8\,\mbox{kpc})^2\big]
\end{equation}
and a total mass $M_\mathrm{bulge}=1.2\times 10^{10}\,M_\odot$, and a single exponential disc with density profile
\begin{equation}  \label{eq:rho_disc}
\rho_\mathrm{disc} \propto \exp\big[-R/3\,\mbox{kpc}\big]\; \cosh^{-2}\big[z/0.5\,\mbox{kpc}\big]
\end{equation}
and a total mass $M_\mathrm{disc}=5\times 10^{10}\,M_\odot$.
The circular velocity at the Solar radius is $\sim 230$~\kms, of which stars contribute $\sim 180$~\kms. 
The bulge and the halo have isotropic velocity distributions by default, but we also created a variant of the Milky Way model with a radially anisotropic halo having $\beta\equiv 1 - \sigma_\mathrm{tan}^2/(2\sigma_\mathrm{rad}^2) = 0.5$.
The two-component equilibrium model for the Milky Way is constructed with the Schwarzschild orbit-superposition method, as implemented in \textsc{Agama}. 

\begin{figure}
\includegraphics{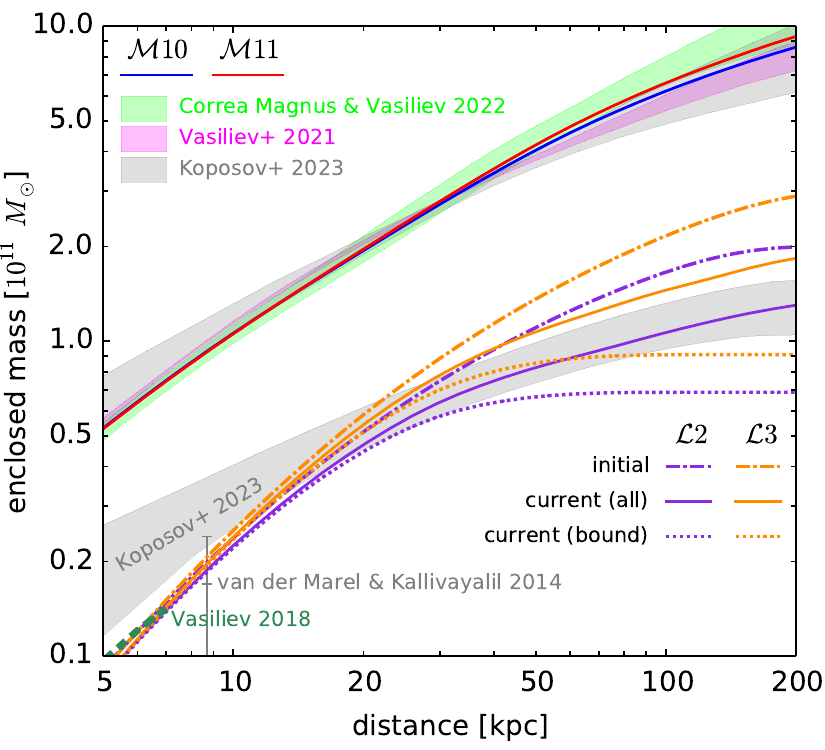}
\caption{
Enclosed mass profiles of the Milky Way (top series of curves) and LMC (bottom series). The two Milky Way models, \M10 and \M11, differ by $\lesssim10\%$ in the plotted radial range and change little during the evolution (shown are the current profiles). By contrast, the LMC models \L2 and \L3 initially differ by a factor of 1.5 (dot-dashed curves), and are significantly reduced in amplitude in the outer regions after the first pericentre passage. Solid curves show the current mass profiles of all LMC particles, and dotted curves -- only the bound particles. Shaded bands show observational constraints from several methods: Milky Way satellites \citep{CorreaMagnus2022}, Sagittarius stream modelling \citep{Vasiliev2021}, Orphan--Chenab stream modelling \citep{Koposov2023} -- all three studies take into account the LMC flyby. The LMC mass profile across all radii is also constrained in the latter study, and the mass in the inner few kpc is measured from stellar kinematics by \citet{vanderMarel2014} and \citet{Vasiliev2018}.
}  \label{fig:massprofiles}
\end{figure}

We consider two variants of models for each galaxy, with the halo parameters listed in Table~\ref{tab:densityparams}. The Milky Way's mass profile, shown in Figure~\ref{fig:massprofiles}, matches contemporary observational constraints (see e.g.\ \citealt{Wang2020} for a compilation of recent estimates), with a virial mass being either $1.0\times10^{12}\,M_\odot$ for the lighter model \M10 or $1.1\times10^{12}\,M_\odot$ for the heavier model \M11. The initial mass of the LMC model is higher than the current estimates (see Figure~1 in \citetalias{Vasiliev2023} for a summary), but as it is tidally stripped on the first pericentre passage, its mass drops down to the acceptable range.

\begin{table}
\caption{Parameters for the initial LMC and Milky Way models in the simulations. 
Halo density for both galaxies follows equation~\ref{eq:rho_halo} with a scale radius $r_\mathrm{s}$, cutoff radius $r_\mathrm{c}$ and total mass $M_\mathrm{total}$. For reference, we also list the virial mass $M_\mathrm{vir}$ and radius $r_\mathrm{vir}$, defined by the condition that the mean density $3M_\mathrm{vir}/(4\pi\,r_\mathrm{vir}^3)$ (including $6.2\times10^{10}\,M_\odot$ in stars, in the case of the Milky Way) is 100 times the critical density of the Universe $\rho_\mathrm{crit}\equiv 3 H_0^2/(8\pi G)$, with $H_0=70$~\kms\,Mpc$^{-1}$.
Radii are given in kpc and masses -- in $10^{11}\,M_\odot$. Last column gives the number of particles.
}  \label{tab:densityparams}
\begin{tabular}{llllllll}
& model & $r_\mathrm{s}$ & $r_\mathrm{c}$ & $r_\mathrm{vir}$ & $M_\mathrm{vir}$ & $M_\mathrm{total}$\!& $N_\mathrm{body}$ \\[1mm]
\raisebox{-2mm}[0mm][0mm]{LMC halo} 
& \L2  & 8.95 & 160.9 & 150 & 1.92 & 2.0 & \raisebox{-2mm}[0mm][0mm]{\!\!$2\times10^6$}\\
& \L3  & 11.7 & 220.6 & 169 & 2.76 & 3.0 \\[1mm]
\raisebox{-2mm}[0mm][0mm]{MW halo} 
& \M10 & 15.0 & 500   & 260 & 10.0 & 11.8 & \raisebox{-2mm}[0mm][0mm]{\!\!$7\times10^6$}\\
& \M11 & 16.5 & 500   & 268 & 11.0 & 12.9 \\[1mm]
\makebox[0cm][l]{MW stars\qquad see equations \ref{eq:rho_bulge} and \ref{eq:rho_disc} in the text} &
& & & & & 0.62 & $10^6$
\end{tabular}
\end{table}

\subsection{Simulation details}  \label{sec:nbodysimulations}

The simulations are conducted using the fast-multipole code \textsc{gyrfalcON} \citep{Dehnen2000}, augmented with a custom plugin for on-the-fly tracing of the centres of both galaxies. We use up to $10^7$ particles in the final highest-resolution runs, but the early stages are run with a reduced number of particles (by factors of 4 or 20). We set the softening length to 0.5~kpc for halo components and 0.2~kpc for the Milky Way stars (the equivalent Plummer softening length is $\sim1.5\times$ smaller), and use a block timestep scheme with the smallest and largest timesteps of 0.5~Myr and 8~Myr, respectively. Since the public version of the code is not parallelized, a single full-resolution simulation takes $\sim 40$ hours.

We adopt the LMC centre position $\alpha=81.28^\circ$, $\delta=-69.78^\circ$ and PM $\mu_{\alpha,\delta}=\{1.858, 0.385\}$~\masyr from \citet{Luri2021}, distance of 49.6 kpc from \citet{Pietrzynski2019}, and heliocentric line-of-sight velocity of 262.5~\kms from \citet{vanderMarel2002}. 
Using a right-handed Galactocentric reference system with the Solar position at \{$-8.12$, 0, 0.02\}~kpc and velocity \{12.9, 245.6, 7.8\}~\kms \citep{Astropy, Drimmel2018}, this translates into the LMC position \{$-0.6$, $-41.0$, $-26.8$\}~kpc and velocity \{$-69.9$, $-221.7$, $214.2$\}~\kms.
Of course, these quantities have non-negligible uncertainties, which have a significant impact on the past orbit reconstruction (see Figure~2 in \citetalias{Vasiliev2023} and associated discussion), but to facilitate the comparison between different simulations, we aim at reproducing the adopted position and velocity as precisely as possible.

The first guess for the initial phase-space coordinates of the LMC $\boldsymbol w_\mathrm{init} \equiv \{\boldsymbol x_\mathrm{init}, \boldsymbol v_\mathrm{init}\}$ may come from integrating its test-particle orbit back in time in the static Milky Way potential, but this is very inaccurate for the massive LMC. A better guess could be obtained by integrating the coupled equations of motion of both galaxies as if they were rigid (non-deforming) but moving points sourcing their respective potentials, adding a drag force responsible for the dynamical friction \citep{Gomez2015,Jethwa2016,Erkal2019}. This approach is adopted in many studies \citep[e.g.,][]{Laporte2018b,GaravitoCamargo2019,Petersen2020}, but since the galaxies in live $N$-body simulations are distorted in the interaction, their orbits deviate from the adopted ones. Previous studies attempted to minimize the deviation by tuning the amplitude of dynamical friction, or by iteratively refining the initial conditions using a grid search in the velocity space and picking up the closest point \citep[e.g.,][]{Donaldson2022,Lilleengen2023}%
\footnote{Instead of a grid search, \citet{Guglielmo2014} used a genetic algorithm to tune the initial conditions, however, their simulations were not fully self-consistent (the Milky Way was modelled as a static potential).}. However, the present-day position and velocity of the LMC $\boldsymbol w_\mathrm{final} \equiv \{\boldsymbol x_\mathrm{final}, \boldsymbol v_\mathrm{final}\}$ are typically still off by a few kpc and few tens of \kms -- an error too large for an accurate reconstruction of the orbit.

In the present study, we aim to radically improve the precision of matching the final phase-space coordinates by iteratively adjusting the initial conditions $\boldsymbol w_\mathrm{init}$. In each iteration, we run a  master simulation with a given $\boldsymbol w_\mathrm{init}$ and a suite of companion simulations with slightly different initial conditions, and determine their final phase-space coordinates. Then we measure the Jacobian of the mapping from $\boldsymbol w_\mathrm{init}$ to $\boldsymbol w_\mathrm{final}$ and use it to assign the next guess for $\boldsymbol w_\mathrm{init}$ (a new value \textit{not} chosen from the current suite). The entire process is then repeated several times until the mismatch in $\boldsymbol w_\mathrm{final}$ drops to an acceptable level ($\lesssim 1$~kpc and 1~\kms). The details of this procedure and associated challenges are described in the \hyperref[sec:appendix]{Appendix}.

The LMC potential represents all particles that originated from the LMC, but many of them are no longer bound to it. To determine the bound mass, we use the following procedure: starting with all LMC particles, we construct their total potential and compute the particle energies $E$ in the reference frame associated with the LMC (i.e., subtracting its centre velocity from particle velocities when computing the kinetic energy). Then particles with $E>0$ are discarded, and the potential is recomputed from the remaining particles, followed by another round of eliminating particles with $E>0$ and repeating the cycle until convergence. The concept of bound mass is ill-defined in the case of an eccentric orbit, because some particles classified as unbound near the pericentre may be subsequently recaptured, but the above procedure gives a reasonable qualitative approximation. The lower set of curves in Figure~\ref{fig:massprofiles} illustrate the enclosed mass profiles for two models, \L2 and \L3, in three variants: initial snapshot, current distribution of all particles originally belonging to the LMC system, and only the currently bound ones. These curves largely coincide within 20 kpc from the LMC centre, but start to deviate further out, with the bound mass essentially reaching a limit around 50 kpc. A comparison with the gray-shaded region representing dynamical constraints on the enclosed mass from the Orphan--Chenab stream modelling \citep{Koposov2023} demonstrates that the two sets of LMC models are on the opposite ends of this range, and likely bracket the actual profile.

\subsection{Resimulating the orbits in the approximated potential}  \label{sec:resimulation}

\begin{figure}
\includegraphics{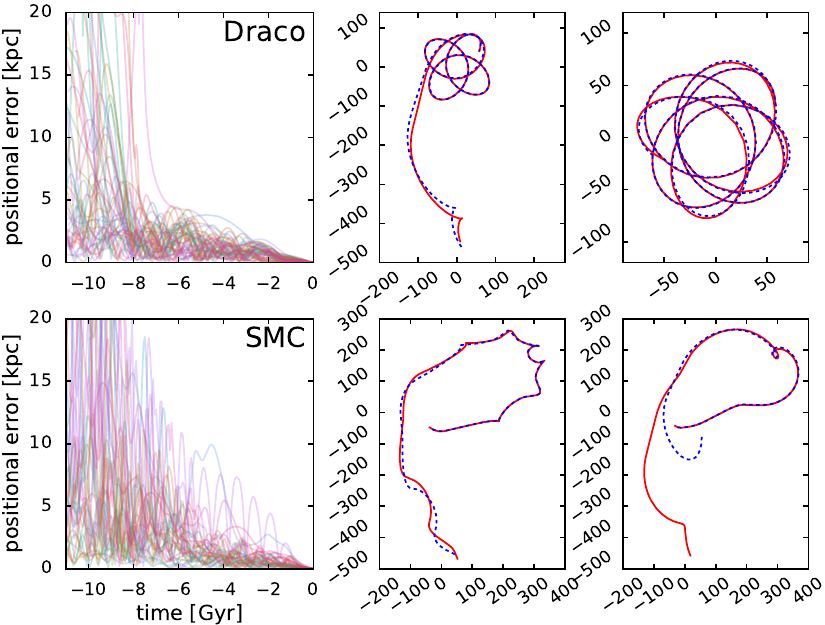}
\caption{Illustration of accuracy of orbit reconstruction, using possible orbits of two satellite galaxies as an example. Left column shows the difference in position between the trajectories of particles in the original simulation and their reconstructed orbits in the approximated potential, starting from the present-day snapshot. Remaining panels show examples of original (red solid) and reconstructed (blue dotted) orbits, which stay fairly close in the majority of cases, although sometimes (less than a few percent cases) may diverge catastrophically (bottom right panel).
}  \label{fig:resim}
\end{figure}

For subsequent analysis, we construct smooth potential approximations for each snapshot in the simulation (stored every 64~Myr), separately for particles belonging to the LMC and to the Milky Way halo. It is convenient to work in the reference frame associated with the Milky Way centre. Because this frame is non-inertial, the total potential contains a spatially uniform but time-dependent \texttt{UniformAcceleration} term, which is obtained from the second derivative of the Milky Way trajectory (it is essential to smooth the trajectory beforehand, as otherwise the second derivative is far too noisy, see section 3.1.2 in \citealt{Sanders2020}). The Galactic stellar disc is approximated by the initial potential at all times, the Galactic halo is represented by a sequence of \texttt{Multipole} potentials pinned to origin, and the LMC is represented by a sequence of \texttt{Multipole} potentials moving along the smooth trajectory extracted from the simulation. We also created alternative \texttt{BasisSet} potential expansions, which use the \citet{Hernquist1992} basis set and are compatible with other dynamical modelling packages such as \textsc{Gala} \citep{PriceWhelan2017}, though these are less efficient to evaluate than the \texttt{Multipole} expansions.

A natural question is whether such a time-dependent potential approximation provides a good description of the actual orbits of particles. \citet{Lowing2011} and \citet{Sanders2020} performed such accuracy analysis for the case of a single host halo from a cosmological simulation, in which there are no LMC-like massive satellites flying around, but numerous subhaloes create small-scale structures in the potential not captured by the low-order basis-set expansion. Nevertheless, it was found that most particle trajectories are well reproduced over the duration of the simulation. In our case, there are no small-scale substructures (apart from the dynamical friction-induced wake in the Milky Way halo), and the LMC is granted its own potential expansion, which moves along a smooth trajectory, so we may expect the fidelity of orbit reconstruction to be even better. Figure~\ref{fig:resim} shows a few examples of orbits from the actual simulation and their reconstructed counterparts, demonstrating that they stay fairly close (to within a few kpc) for the entire duration of simulation in most cases. 
Even if individual orbits may not be perfectly reconstructed, the ensemble of nearby orbits corresponding to a given satellite (Section~\ref{sec:satellite_membership}) is very well reconstructed in the statistical sense. Re-running the entire simulation as an ensemble of non-interacting particles closely reproduces the current distribution and kinematics of Magellanic debris and the corresponding perturbations in the Milky Way halo.

After experimenting with different orders of expansion, we find that $\ell_\mathrm{max}=6$ is more than sufficient, and we use 25 log-spaced nodes between 0.1 and 1000 kpc in the radial grids.
\citet{Sanders2020} also compared the accuracy of orbit approximation in the case when the potential was taken from the nearest snapshot or linearly interpolated between snapshots, and found that the latter makes little difference but is twice more expensive (unless the expansion coefficients themselves are linearly interpolated before computing the potential, which is not available in \textsc{Agama}). However, we noticed that since the orbit integration in \textsc{Agama} uses a high-order method (\textsc{dop853}), the extra cost of a double potential evaluation is more than compensated by its increased smoothness, allowing the integrator to take larger timesteps. Even so, the computational cost (number of timesteps) scales with the requested accuracy $\epsilon$ as $\epsilon^{-1/n}$ with $n\simeq 4$ for the nearest-timestamp potential and $n\simeq 6$ for the linearly interpolated potential, faster than the $n=8$ scaling expected for this integrator in an infinitely smooth potential. Nevertheless, a value of $\epsilon=10^{-4}$, while much coarser than the default $\epsilon=10^{-8}$ used for analytic potentials, is sufficient for the relatively short duration of orbit integration (at most a few tens dynamical times).

%%%%%%%%
\section{Analysis}  \label{sec:analysis}

\subsection{Orbital history of the LMC}  \label{sec:orbit}

\begin{figure}
\includegraphics{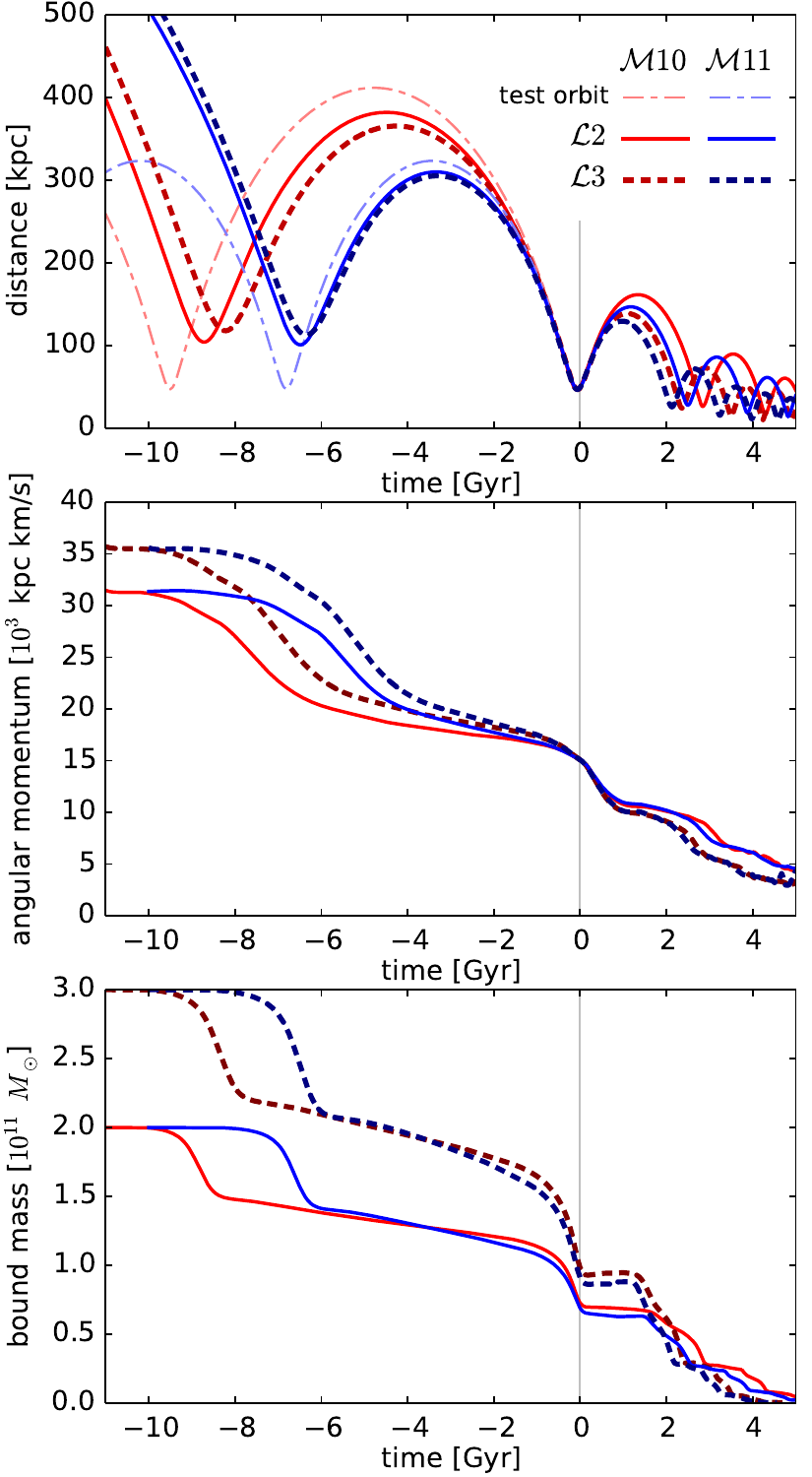}
\caption{
Evolution of the distance between the LMC and the Milky Way (top panel), orbital angular momentum (middle panel) and bound mass of the LMC (bottom panel) for the simulations with different LMC masses and Galactic potentials. In the top panel, we also show the test-particle orbits in the given potential by thin dot-dashed lines. As the LMC mass increases, its inferred previous orbital period and previous apocentre radius are reduced. A less than $10\%$ variation in the Galactic mass between \M10 and \M11 changes the orbital period by $\gtrsim 30\%$. Although both bound mass and angular momentum decrease in sharp steps around pericentre passages, the mass loss occurs predominantly \textit{before} the pericentre, while the angular momentum starts to drop \textit{after} the pericentre (in contrast to the classical dynamical friction scenario, in which it would decrease more symmetrically on both approaching and receding segments of the orbit).
}  \label{fig:orbit}
\end{figure}

\begin{figure*}
\includegraphics{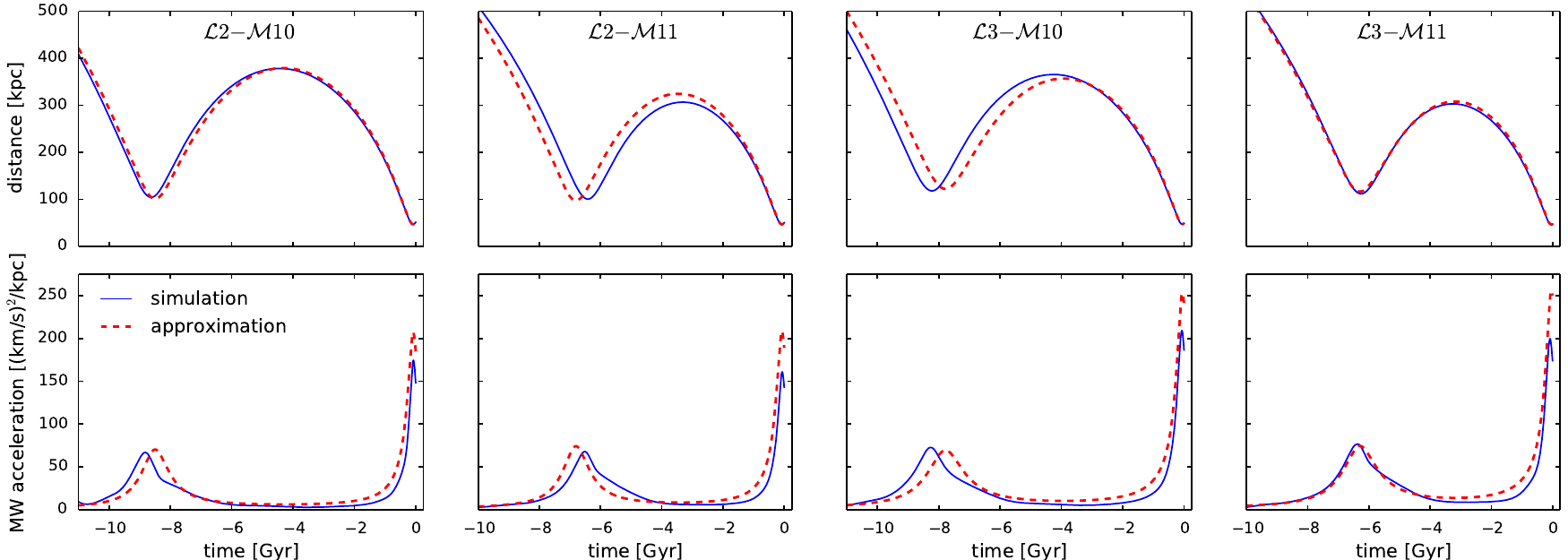}
\caption{
Comparison of past LMC trajectories (top row) and LMC-induced accelerations of the Milky Way (bottom row) between the $N$-body simulations (solid blue lines) and solutions of an approximate ODE system (dashed red lines). Although the latter qualitatively describes the evolution after tuning some of its parameters, there are noticeable deviations in the orbital period and the amplitude of acceleration.
}  \label{fig:orbit_approx}
\end{figure*}

We first examine the past orbits of the LMC for different combinations of its mass and Galactic potential. Figure~\ref{fig:orbit}, top panel, shows the Galactocentric distance of the LMC in four simulations, together with test-particle orbits in the two Milky Way potentials \M10, \M11, shown by dot-dashed curves. It confirms the trend reported in \citetalias{Vasiliev2023}: more massive LMC have smaller apocentre distances and shorter periods. This results from the interplay of two opposite effects. On the one hand, a more massive LMC experiences stronger dynamical friction, thus it must have started with a higher orbital energy to arrive to the same present-day phase-space point. On the other hand, though, a more massive LMC induces a larger reflex motion in the Milky Way, thus to achieve the same measured relative velocity, it needs to have a lower velocity in the centre-of-mass frame, meaning lower total energy and shorter period. Equivalently, the orbital period of the binary system depends on its total mass and is shorter for higher-mass LMC models. It turns out that the second factor outweighs the first one, though the difference in orbital period is rather minor between the two LMC models.

By contrast, the dependence of the orbital period on the Milky Way potential is quite dramatic, with $\gtrsim 30\%$ difference between \M10 and \M11 (which differ by 10\% in virial mass). As discussed in Section~3.2 of \citetalias{Vasiliev2023}, this is due to the orbit being only marginally bound and the LMC currently located close to its pericentre, therefore its kinetic energy is almost equal in magnitude to its potential energy, and even a small variation in the latter leads to dramatic changes in the total energy. It is certainly possible to make the period comparable or longer than the Hubble time (i.e., put the LMC onto a first-approach trajectory), but the period could also be as short as 5--6~Gyr for a reasonable choice of Galactic potential. In these idealised simulations, we neglected the cosmological evolution of the Milky Way mass profile, which would have lengthened the orbital period (see Figure~11 in \citealt{Kallivayalil2013}), but also neglected other factors that could shorten it -- the perturbation of the LMC velocity by the SMC or a favourable orientation of a non-spherical Milky Way potential (both discussed in \citetalias{Vasiliev2023}). Overall, it is plausible that the previous passage of the LMC could have occurred between 5 and 10 Gyr ago at a distance just over 100~kpc, although the details may differ from our setup.

The middle and lower panels show the evolution of angular momentum and bound mass in these four simulations. Both quantities experience stepwise drops around each pericentre passage, but interestingly, the decrease of angular momentum occurs mostly after the pericentre, whereas the decrease of bound mass mostly precedes it. The bound mass during the past orbital period stays in the range (1.2--2)${}\times 10^{11}\,M_\odot$, compatible with the current estimates (see Figure~1 in \citetalias{Vasiliev2023} for a compilation of literature values).

A popular approach for reconstructing the past orbit of the LMC, taking into account dynamical friction from the Galactic halo and reflex motion that the LMC induces on the Milky Way, is the solution of a coupled ODE system for two extended masses (Equation~1 in \citetalias{Vasiliev2023}), using the gradient of gravitational potential of each galaxy at the current distance $r(t)$ to compute the accelerations of both galaxies. It has been followed in many previous studies, e.g., \citet{Gomez2015}, \citet{Jethwa2016}, \citet{Erkal2019}, \citet{Patel2020}, \citet{CorreaMagnus2022}. As the Coulomb logarithm is generally treated as a tunable parameter, \citet{Hashimoto2003} found that a distance-dependent value $\ln\Lambda = \ln(r/r_\mathrm{min})$ provides a better match the results of actual $N$-body simulations, with $r_\mathrm{min}$ comparable to the scale radius $r_\mathrm{s}$ of the satellite. The complication usually ignored by these studies is that the LMC mass also decreases with time, especially in the second-passage scenario. Figure~\ref{fig:orbit_approx} compares the actual orbits in our $N$-body simulations (solid blue) with the solutions of the approximate ODE system (dashed red), in which the LMC mass has been set to half its initial value and $r_\mathrm{min}$ set to $0.8\,r_\mathrm{s}$. There are noticeable deviations in the reconstructed orbital periods (top row), and the LMC-induced acceleration of the Milky Way is overestimated by $\sim$20\% at its peak (bottom row), because this prescription ignores the deformation of both galaxies. Although with two free parameters, we managed to match the grid of 4 simulations reasonably well, there is no guarantee that this recipe would be adequate for all reasonable choices of Milky Way and LMC mass profiles.
It is clear that a more accurate approximation is highly desirable for a fast and reliable quantitative analysis, short of running the actual simulations.

\subsection{Perturbations of the Milky Way halo}  \label{sec:mwhalo}

\begin{figure}
\includegraphics{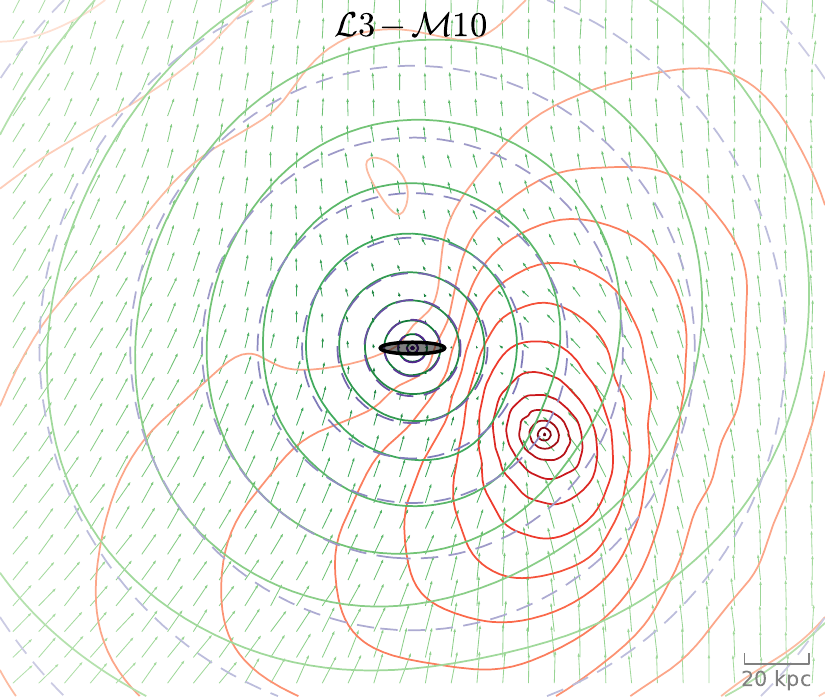}
\includegraphics{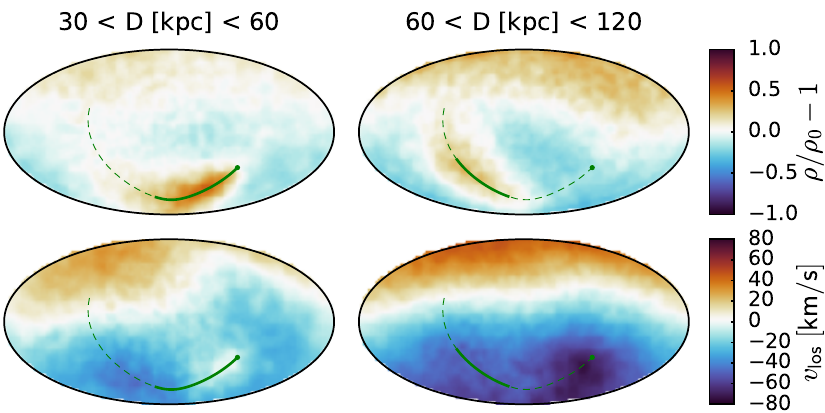}
\caption{
Top panel: projected density of Milky Way (green) and LMC (red) in the orbital plane of the latter, for the simulation \L3--\M11. Dashed blue contours show the projected density of a sphericalized Milky Way halo, thus the deviations between green and blue contours illustrate the distortions (e.g., since the green contours are further out in the top right corner, this is perceived as an overdensity in that direction). The Galactic disc is nearly perpendicular to this plane and its projection is shown by a gray ellipse. Arrows indicate the mean velocity of Milky Way halo particles relative to its centre: the innermost 20--30 kpc have little net velocity, while in the outer regions it is directed primarily towards the North Galactic pole.\protect\\
Middle row: heliocentric view of the density contrast in the Galactic halo. Left and right panels show the intermediate (30--60 kpc) and distant (60--120 kpc) regions. The past orbit of the LMC is shown by a solid green line when within this radial range, or by a dashed line otherwise, and its current location is indicated by a dot. Apart from the dynamical friction-induced overdensity along the orbit, there is a global dipole asymmetry primarily in the outer halo, corresponding to the downward shift of the Milky Way centre relative to its outer regions.\protect\\
Bottom row: heliocentric view of the line-of-sight velocity of the Galactic halo in the same two radial ranges. The north--south dipole asymmetry is most prominent in the outer region.
}  \label{fig:mwhalo}
\end{figure}

\begin{figure}
\includegraphics{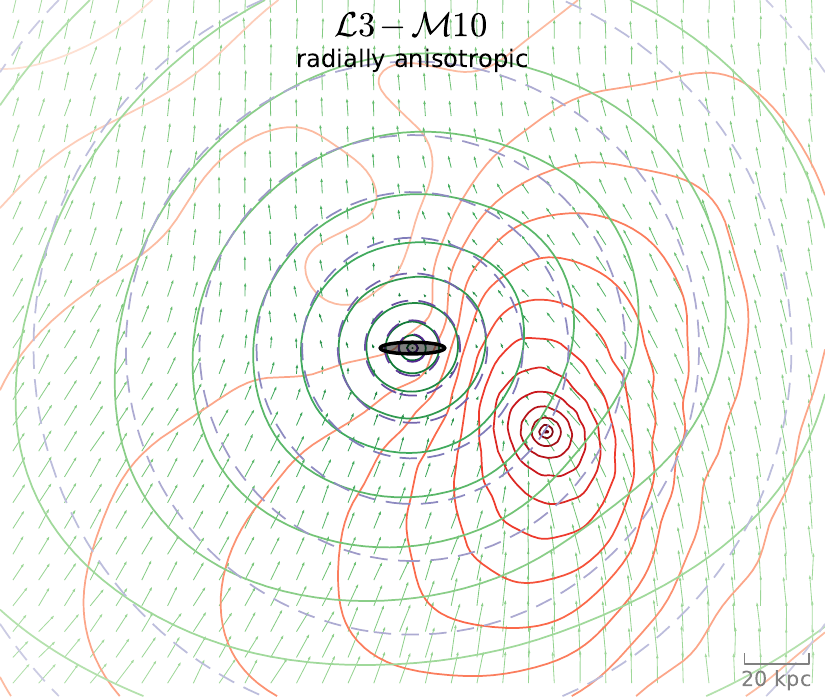}
\includegraphics{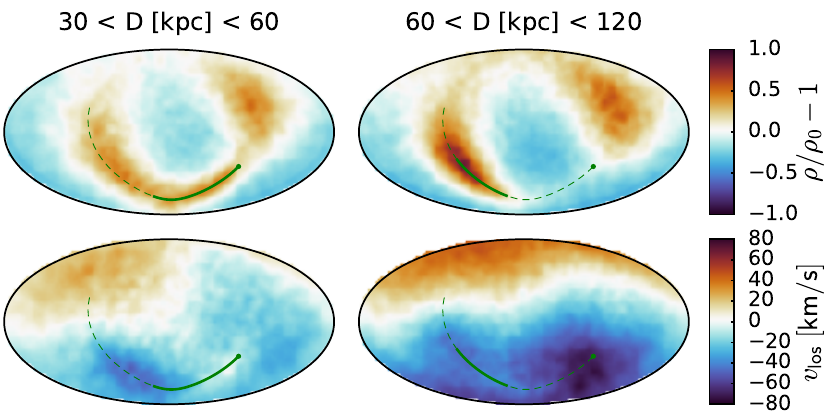}
\caption{
Same as Figure~\ref{fig:mwhalo}, but for the simulation with a radially anisotropic Galactic halo. The most obvious difference is an increased density contrast in the wake and an overdensity ahead of the LMC on its orbit, while the kinematic signatures are largely the same.
}  \label{fig:mwhalo_rad}
\end{figure}

As established in several recent studies \citep{GaravitoCamargo2019,GaravitoCamargo2021a,Cunningham2020,Erkal2020b,Erkal2021,Petersen2020,Petersen2021}, a massive LMC induces significant perturbations in the Galactic halo at distances $\gtrsim 20$--30~kpc (see Section~4.2 in \citetalias{Vasiliev2023} for an extended discussion). A natural question is, therefore, whether the second-passage scenario produces different signatures from the first-passage scenario considered in the above papers.

Figure~\ref{fig:mwhalo} shows the contours of Milky Way halo density projected onto the LMC orbital plane (top panel) and heliocentric sky maps of density and line-of-sight velocity perturbations (middle and bottom rows) in the \L3--\M10 simulation. At a glance, these plots look very similar to the ones from the first-passage simulations (e.g., Figure~5 in \citetalias{Vasiliev2023}). For a more detailed analysis, we constructed another simulation of a first passage, using the same Galactic model \M10 and a lower-mass LMC model \L2, started near the apocentre at time 4~Gyr ago. Of course, such a trajectory extrapolated further into the past would have had another pericentre passage, but we ignore this inconsistency for the sake of experiment. The density and kinematic perturbation maps in this scenario look so similar to Figure~\ref{fig:mwhalo} that it is not even worth showing them separately. In other words, any possible signatures of the first passage either dissipate or are totally obscured by the most recent passage, which occurred at a twice smaller distance.

Models with the same LMC mass but different Milky Way potentials again look quite similar, and the variation of LMC mass only changes the amplitude of perturbations, but not their qualitative appearance. On the other hand, there are interesting differences between the models with isotropic Milky Way halo discussed so far, and a model with a radially anisotropic halo ($\beta = 0.5$, typical for the outer region of NFW haloes, e.g., \citealt{Hansen2006}), which is shown in Figure~\ref{fig:mwhalo_rad}. The velocity perturbations remain largely the same as in the previous figure, but the dynamical friction-induced overdensity along the past orbit of the LMC is much more prominent, and even more intriguingly, there is another overdensity along the future trajectory (upper left corner of the sky map). The large-scale dipole asymmetry is not significantly affected, but the local wake is strongly enhanced, in complete agreement with the study of \citet{Rozier2022}, who used a very different method (linear response theory instead of conventional $N$-body simulations). It is unclear which of the two cases better matches the current observations: a density asymmetry is much more difficult to measure reliably than a kinematic perturbation, and the only study which attempted this \citep{Conroy2021} does not match either plot in detail (although they do note that the measured density contrast in the local wake is significantly higher than even in their simulation with the highest LMC mass of $2.5\times10^{11}\,M_\odot$). The previous orbital period of the LMC in the radially anisotropic halo is similar to its isotropic counterpart, but the initial angular momentum is higher and the distance of the first pericentre is correspondingly larger by $\sim 20\%$; in other words, radial anisotropy of the host galaxy enhances the radialization of satellite orbits, as already mentioned in \citet{Vasiliev2022}.

It is also worth noting that the observationally determined direction of the dipole perturbation in line-of-sight velocities \citep{Petersen2021} differs by more than 45$^\circ$ from the roughly north--south direction found in all simulations, including those in the present paper; the reasons for this discrepancy are not clear and may point to some deficiencies of current models, first or second passage alike.

\subsection{Perturbations of the Galactic disc}  \label{sec:mwdisc}

\begin{figure}
\includegraphics{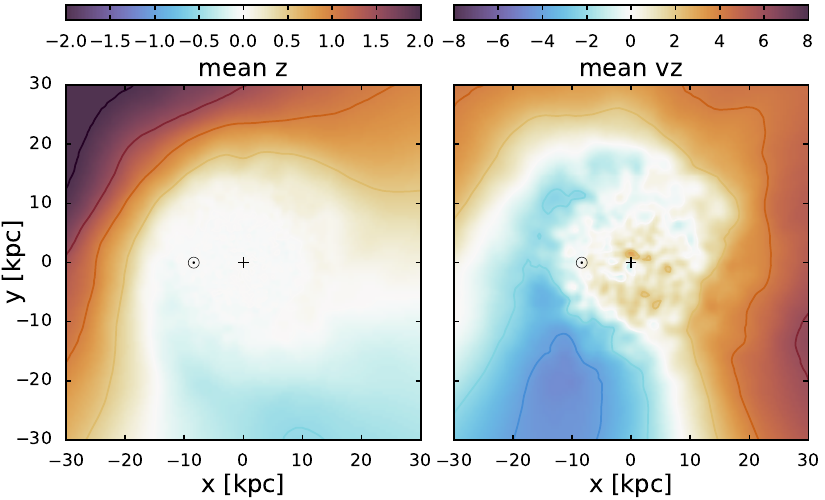}
\caption{
Effect of the LMC on the Milky Way disc in the \L3--\M11 simulation. Left panel shows the mean displacement of disc particles from the $z'=0$ plane, where the axis $z'$ is already tilted by 1.5$^\circ$ from the initial direction $z$ to better align the inner 10~kpc of the disc with the $z'=0$ plane. Right panel shows the mean vertical velocity along this tilted axis. The Galactic centre is marked by \textbf{+} and the Solar position -- by $\odot$.
}  \label{fig:warp}
\end{figure}

The LMC also imparts a noticeable perturbation on the Milky Way disc, likely contributing to the warp observed in its outer regions both in stars and gas.
One should keep in mind that the very definition of the disc plane is not unambiguous. Here we use the particles within 10~kpc from the Galactic centre to construct a plane that minimizes the scatter of their $z'$ coordinates relative to that plane. Although the initial models had $z'$ aligned with the $z$ axis, by $t=0$ this best-fitting plane is already tilted by 1--1.5$^\circ$, which has an unintended consequence that the LMC coordinates in this rotated system are slightly different from the input values. Fortunately, the tilt direction is such that this offset roughly corresponds to a shift along the same trajectory by $\sim$3--4~Myr, so we ignore this minor deviation instead of trying to fit its present-day position in the rotated coordinate system. 

Figure~\ref{fig:warp} shows the mean $z$ coordinate ($z_0$) of stars across the Galactic disc in the \L3--\M11 simulation, which is qualitatively similar to other simulations. The noticeable warp in the top left corner is nearly opposite to the current location of the LMC ($x\approx -1, y\approx -41$~kpc), and is caused by the same effect as the distortions in the Galactic halo, namely, that the inner part of the disc has shifted downward relative to the regions further away from the LMC. This displacement map qualitatively resembles Figure~6 in \citet{Laporte2018a}, who conducted a similar set of simulations, though without matching the present-day LMC coordinates as precisely as in the present study. We stress that the perturbations rapidly evolve, e.g., the direction of the warp changes by $\sim\!90^\circ$ over 0.1~Gyr, so for a quantitative analysis and comparison with observations, it is important to have a precisely fitted model. Here we do not attempt a detailed comparison, only noting that the warp in the model is similar to the one observed in stars (e.g., Figure~7 in \citealt{RomeroGomez2019}, or Figure~2 in \citealt{Lemasle2022}), but somewhat smaller in amplitude. \citet{Laporte2018b} found that a combination of perturbations from the LMC and Sagittarius may create a stronger warp signature than a simple superposition, thus we do not expect our LMC-only models to match the data perfectly.

\subsection{Future evolution}  \label{sec:future}

\begin{figure}
\includegraphics{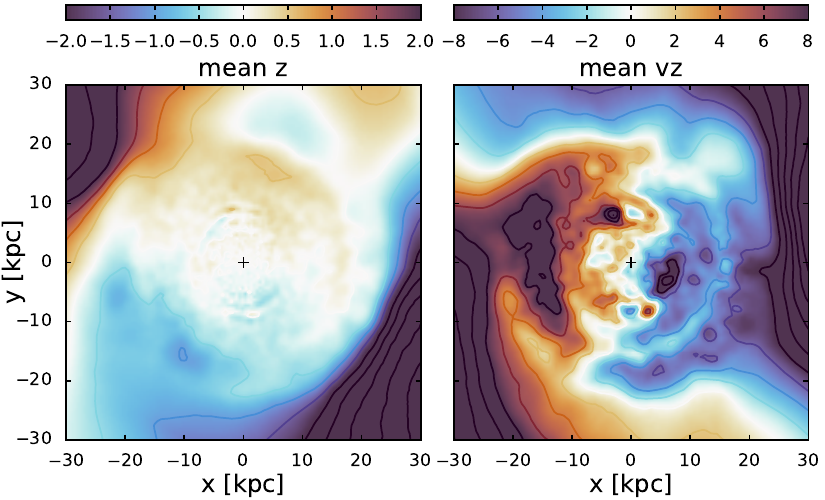}
\includegraphics{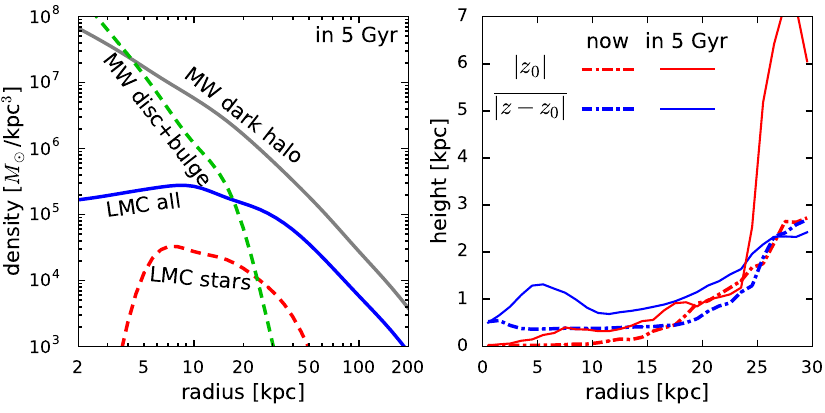}
\caption{
Effect of the LMC on the Milky Way disc in the future (5 Gyr from now), when the LMC is fully disrupted. Top row shows the maps of mean vertical displacement and vertical velocity relative to the $z'=0$ plane, where the $z'$ axis is tilted by 10$^\circ$ from the initial direction $z$; compared to Figure~\ref{fig:warp}, the perturbations are much stronger. 
Lower left panel shows the density of different subsets of particles at $t=5$~Gyr: Milky Way dark halo (gray solid), Milky Way stars (green dashed), LMC particles (blue solid), and the 1\% initially most bound LMC particles approximating its stellar component (red dashed).
Lower right panel shows the warp amplitude $|z_0|$ (red) and disc thickness $\overline{|z-z_0|}$ (blue) at present time (dot-dashed) and in 5 Gyr (solid). The disc thickness will nearly double after the LMC impact, and the outer region beyond 25 kpc will be strongly displaced from the $z'=0$ plane.
}  \label{fig:future}
\end{figure}

Given that our simulations produce an LMC model not only matching its observed present-day position and velocity, but also dynamically consistent with its prior evolution (i.e., with appropriate degree of tidal stripping), it makes sense to explore the future fate of the LMC and its effect on the Milky Way. We run the simulations for another 5 Gyr, and depending on the LMC mass, it completes another 3--4 orbits before being entirely disrupted (although the lowest-mass combination, \L2--\M10, is not fully merged by the end of this interval). Figure~\ref{fig:orbit} shows the future evolution of its mass and Galactocentric distance to the right of the vertical gray line. 

Although the current magnitude of perturbations in the Galactic halo will not be exceeded in future pericentre passages due to the rapidly decreasing LMC mass, its effect on the Galactic disc will be more prominent as the pericentre radius drops to $\sim$25~kpc on the next passage, and then further to 10--15~kpc. Figure~\ref{fig:future}, top row, shows the maps of the disc warp similar to Figure~\ref{fig:warp}, but at 5 Gyr in the future for the most massive combination \L3--\M11, in which the LMC is fully disrupted by that time. The perturbations in the disc are much stronger than now, and it has become significantly hotter (lower right panel shows that the thickness has doubled across all radii) and tilted by $\gtrsim 10^\circ$ relative to its initial orientation, but not destroyed.

Lower left panel of that figure shows the density profile of Magellanic debris, which has two characteristic break radii at 10~kpc (the radius of the last pericentre) and 40--50~kpc (the current pericentre radius). Although the initial LMC models did not have a separate stellar component, one could get a crude but sensible approximation by tracking 1\% of most bound particles (the mass $3\times10^9\,M_\odot$ roughly matches the estimate of total stellar mass from \citealt{vanderMarel2002}). The density of these particles is shown by red dashed line, and stays well below the density of Milky Way disc (dashed green) in the inner Galaxy, but may be comparable or even exceed the density of the pre-existing stellar halo at large distances. The total mass of present-day Milky Way stellar halo is estimated to be $\sim 10^9\,M_\odot$ \citep{Deason2019}, so the LMC will quadruple it when fully disrupted. These findings agree well with the predictions of \citet{Cautun2019} based on Milky Way--LMC analogues in the EAGLE simulation, although in our experiments, it takes somewhat longer (4--5~Gyr) to fully destroy the LMC.

%%%%%%%%
\section{Galactic and Magellanic satellites}  \label{sec:satellites}

We now turn to the analysis of the population of dwarf galaxies surrounding the Milky Way (colloquially called Galactic satellites, although some of them are in fact satellites of the LMC).

\subsection{Previous work}  \label{sec:satellites_literature}

\citet{LyndenBell1976} noticed that the dwarf galaxies Draco, Ursa Minor and Sculptor lie close to the plane of the Magellanic gas stream, itself discovered shortly before \citep{Mathewson1974}, and proposed that they are tidally stripped satellites of the postulated Great Magellanic Galaxy -- the precursor of the LMC itself. \citet{LyndenBell1982a} added Carina dSph to the list of galaxies possibly associated with the Great Magellanic Galaxy, while \citet{LyndenBell1982b} introduced another dSph grouping -- Fornax, Leo~I, Leo~II and Sculptor (FLS), but treated them separately from the Magellanic group, and \citet{Majewski1994} added Sextans and Phoenix dSph to the FLS group. However, the planes of two groups differ by only $30^\circ$, so \citet{Kroupa2005} proposed that they form a single group, later dubbed ``Vast Polar Structure'' by \citet{Pawlowski2012}. Although a spatial alignment could happen by chance, it gradually became clear that most of these satellites also have similar directions of angular momentum (orbital poles), making this structure dynamically coherent \citep{Metz2008,Pawlowski2013,Pawlowski2020,Fritz2018}.

\citet{DOnghia2008} proposed a Magellanic group consisting of 7 galaxies (two Clouds, Draco, Leo~II, Sagittarius, Sextans and Ursa Minor), although they did not comment on why the orbital plane of Sagittarius is nearly orthogonal to that of the LMC. \citet{Metz2008} presented the now canonical list of clustered orbital poles of classical dwarfs: the Clouds, Carina, Draco, Fornax, Ursa Minor, and noted that if the Clouds were on their first passage and unrelated to the other four galaxies that have been orbiting the Milky Way for a while, it would be a rather unlikely coincidence for them to share the same orbital plane just by chance. 
Although the idea that a relatively thin and kinematically coherent disc of satellites could result from an accretion of a massive group of galaxies early in the Milky Way history is quite natural \citep[e.g.,][]{Li2008,Smith2016}, somehow it did not get much traction, with \citet{Metz2009} arguing against it on the grounds that the observed plane of satellites in the Milky Way is too compact compared to the expectations from group accretion. Instead, \citet{Pawlowski2011} proposed a scenario in which the VPOS consists of ``tidal dwarf galaxies'' formed out of gas expelled during a tidal interaction between the Milky Way and another galaxy early in its evolution, somewhat reminiscent of the now abandoned Jeans--Jeffreys tidal theory of the Solar system formation \citep{Jeffreys1952}. They considered the possibility that the proto-LMC could be this intruder galaxy, though in their scenario the pericentre passage leading to the tidal stripping would have occurred at a much smaller distance ($\sim 10$~kpc) than the most recent pericentre distance of the LMC (45--50~kpc).

\begin{table}
\caption{A selection of possible associations of classical dSph with the Magellanic group according to different studies:
\citet[L76]{LyndenBell1976}, \citet[D08]{DOnghia2008}, \citet[M08]{Metz2008}, \citet[N11]{Nichols2011}, \citet[J19]{Jahn2019}, \citet[P20]{Pardy2020}, \citet[E20]{Erkal2020a}, \citet[S21]{SantosSantos2021}, \citet[B22]{Battaglia2022}. Plus, minus and question mark indicate likely, unlikely and uncertain association (not necessarily that the galaxy is currently a member of the Magellanic system), while objects not discussed in a particular paper are shown by empty spaces; SMC is not listed because it is included as a member (sometimes implicitly) in all studies. A complementary Table~1 in \citetalias{Vasiliev2023} lists studies that considered more recently discovered ultrafaint dwarfs and focuses on the current satellites of the LMC.
}  \label{tab:satellites}
\begin{tabular}{lccccccccc}
galaxy &\!\!L76\!\!&\!\!D08\!\!&\!\!M08\!\!&\!\!N11\!\!&\!\!J19\!\!&\!\!P20\!\!&\!\!E20\!\!&\!\!S21\!\!&\!\!B22\!\!\\[1mm]
%           L76 D08 M08 N11 J19 P20 E20 S21 B22
Carina     & ? &   & + &-- & + & + &-- & ? & ? \\
Draco      & + & + & + & + &   &   &-- &   &   \\
Fornax     &   &   & + & ? & + & + & ? & ? &-- \\
Leo~I      &   &   &   &-- &   &   &-- &   &   \\
Leo~II     &   & + &   &   &   &   &-- &   &   \\
Sagittarius&   & + &-- & + &   &   &-- &   &   \\
Sculptor   & + &   &-- & + &   &   &-- &   &   \\
Sextans    &   & + &   & ? &   &   &-- &   &   \\
Ursa Minor & + & + & + & + &   &   &-- &   &   \\
\end{tabular}
\end{table}

Until about 10 years ago, the census of Milky Way satellites was limited to relatively bright objects, whose association with the Magellanic system is difficult to establish with confidence (apart from the aforementioned similarity of orbital pole directions). Table~\ref{tab:satellites} summarizes the proposed claims of Magellanic origin for classical dwarfs. However, since then many new ultrafaint galaxies were discovered in deep photometric surveys such as DES, and quite a few of them are very likely to be LMC satellites (see Table~1 in \citetalias{Vasiliev2023} for a compilation of results).

Numerous studies that considered the association of dSph and ultrafaint galaxies with the LMC system can be broadly divided into several categories:
\begin{enumerate}
\item One may take the present-day positions and velocities of these objects as well as those of the LMC itself, then reconstruct the past orbit of the LMC in the assumed Galactic potential (ideally, not as a test particle moving in a fixed potential, but taking into account the reflex motion it induces on the Milky Way), and then follow the orbits of dwarf galaxies in the combined potential of the Milky Way and the LMC. An object can be declared an LMC satellite if it is currently bound to it, or a former satellite if it was bound at some point in the past (i.e., its relative velocity w.r.t.\ the LMC was below the escape velocity of the latter). Examples of this approach include \citet{Erkal2020a}, \citet{Patel2020}, \citet{Battaglia2022}, \citet{CorreaMagnus2022}, \citet{Pace2022}. The limitation of this approach is that it needs high-accuracy 6d phase-space coordinates of all objects for a reliable reconstruction of past orbits, which are rarely available, and even then the reconstruction becomes increasingly less reliable further into the past \citep[e.g.,][]{DSouza2022}. This creates a bias against the association with the LMC system, since it is much more likely that the measurement errors will scatter an orbit away from the LMC than closer to it.
\item To combat the latter problem, one may instead perform forward orbit modelling, namely, consider a large sample of particles initially bound to the LMC system, evolve their orbits forward up to the present time, and then identify which of the observed satellites are close to any of these LMC-associated particles within uncertainties. This alone is not enough to establish the likelihood of association -- one needs to compare it with the alternative hypothesis that an object comes from the Milky Way system (or was accreted onto it independently from the LMC). This could be achieved by sampling an equally large number of particles from the Milky Way halo and counting the fraction of particles in the spatial and kinematic neighbourhood of each galaxy coming from the two alternative sources. The first part of this approach was followed in \citet{Nichols2011}, \citet{Yozin2015}, \citet{Jethwa2016}, but these studies only examined whether a given dSph lies in the region occupied by the LMC debris (not even comparing the velocities due to a lack of measurements) without considering the alternative (non-LMC) origin.
\item One can take this idea further and perform a full $N$-body simulation of the two interacting galaxies, comparing the positions and velocities of observed dSph with the spatial and kinematic distribution of particles originating in both galaxies. Several groups have conducted simulations of this kind \citep{GaravitoCamargo2019, Petersen2020, Vasiliev2021, Pawlowski2022, Makarov2023}, but none so far have used them to assess the satellite population of the Magellanic system, although \citet{Vasiliev2021} employed this approach to determine the association of globular clusters with the Sagittarius galaxy (which was also simulated as a live $N$-body system in that study). The challenge is to match the current position and velocity of the LMC with sufficient precision for this analysis to be meaningful, and only the latter study took great effort to meet it.
\item Finally, one can resort to the analysis of cosmological simulations, either in large volume with thousands of Milky Way analogues, such as Millenium-II \citep{BoylanKolchin2011}, or more tailored Local Group-scale simulations, such as Aquarius \citep{Sales2011,Sales2017,Kallivayalil2018}, ELVIS \citep{Deason2015}, FIRE \citep{Jahn2019}, Auriga \citep{Pardy2020} or Apostle \citep{SantosSantos2021}. In this case the main challenge is to find suitable analogues of the Milky Way--LMC system, abandoning any hope for a detailed match, but only qualitatively reproducing its main properties. Earlier studies, such as \citet{Sales2011}, did not find conclusive evidence of Magellanic association for any of the then-known dSph galaxies (except the SMC, of course), but with the more recent discoveries of numerous ultrafaint dwarfs in the vicinity of the LMC, several of them were proposed to be its satellites based on the similarity of their location (and sometimes velocity) to particles from LMC analogues in the simulations. \citet{Jahn2019} and \citet{Pardy2020} also proposed Carina and Fornax as possible former satellites of the LMC, based on their updated PM from \Gaia DR2, but \citet{SantosSantos2021} found this possibility rather tenuous.
\end{enumerate}

In this study, we take the third approach, analyzing the possible origin of each dSph by comparing its phase-space coordinates with particles in the simulation. We also use elements of the second approach, rewinding dSph orbits in the pre-recorded potential of the simulation approximated by smooth potential expansions and comparing the probability of arriving with the Magellanic system against originating in the Milky Way. We take the coordinates, distances and line-of-sight velocities from the catalogue of \citet[updated in 2021]{McConnachie2012}\footnote{\url{https://www.cadc-ccda.hia-iha.nrc-cnrc.gc.ca/en/community/nearby/}}, and \Gaia eDR3 PM from \citet{Battaglia2022}; the SMC PM is taken from \citet{Luri2021}.

\subsection{Satellites in our simulations}  \label{sec:satellite_membership}

Let $\boldsymbol X$ be the 6d vector of observable phase-space coordinates of a given object (longitude $l$, latitude $b$, distance $D$, two PM components $\mu_l$, $\mu_b$, and line-of-sight velocity $V$)\footnote{Here longitude and latitude are not in the Galactic coordinate system, but rather in a system rotated such that the target object is at origin, minimizing distortions of the PM field that could appear for objects near celestial poles.}, and $\mathsf E$ be the covariance matrix of measurement uncertainties (in case of uncorrelated errors, it is diagonal with squared standard deviations $\epsilon_c$ in each coordinate $c$). Although the position is measured essentially perfectly, we need to introduce an artificial uncertainty in celestial coordinates in order to have a sufficient number of particles to match, setting $\epsilon_{l,b}=4^\circ$. We also put a lower limit on the uncertainty in distance ($\epsilon_D \ge 0.1\,D$), line-of-sight velocity ($\epsilon_V \ge 10\,$\kms) and PM ($\epsilon_\mu \ge 0.05\,$\masyr), both to increase the chance of matching and to account for various systematic deviations between the model and the real Milky Way.
The $N$-body snapshot is then converted into the same coordinates $\boldsymbol{X}^{(i)}$, $i=1..N_\mathrm{body}$. To determine the probability of the object to belong to either the LMC or the Milky Way, we use three different approaches.

In the first method (particle matching), we compute the uncertainty-weighted Mahalanobis distance between the object and each $N$-body particle: $d^{(i)} \equiv \sqrt{ \big(\boldsymbol X - \boldsymbol X^{(i)}\big)^T\; \mathsf{E}^{-1}\; \big(\boldsymbol X - \boldsymbol X^{(i)}\big) }$. The matching likelihood for $i$-th particle is $p^{(i)} \propto m^{(i)}\,\exp\big[-\frac{1}{2}(d^{(i)})^2 \big]$, where $m^{(i)}$ is the mass of this particle, and the overall coefficient of proportionality determined from the condition that $\sum_{i=1}^{N_\mathrm{body}} p^{(i)} = 1$. Then the probability of association with the LMC is simply the sum of $p^{(i)}$ over particles that originally belonged to the LMC. One can obtain a representative sample of past orbits by selecting particles in proportion to their likelihood of matching and retrieving their trajectories from the original simulation. This method is the most direct, but may suffer from low-number statistics even in the highest-resolution snapshots, if the number of particles with sufficiently high matching probabilities is too small. In these cases, we artificially increase the uncertainties on kinematic measurements ($\mu_{l,b}$ and $V$) until the highest matching probability is below 0.1 (i.e., there are at least 10 different matches).

\begin{figure}
\includegraphics{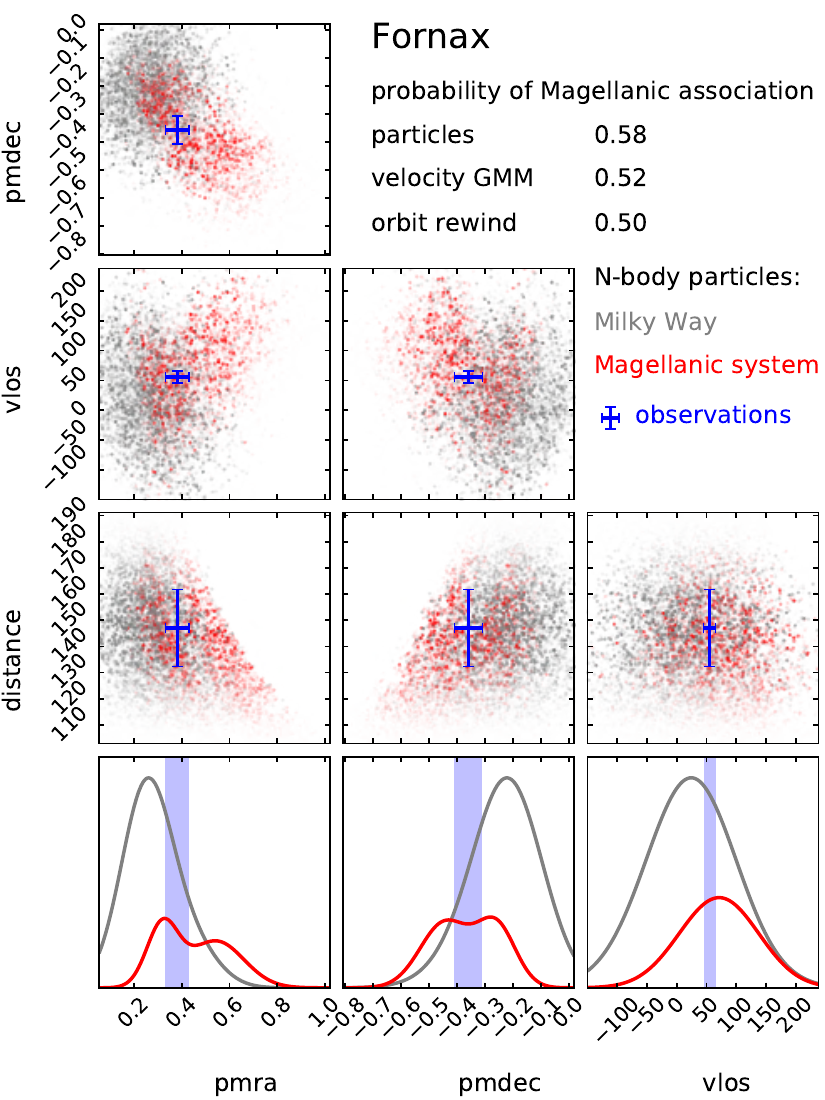}
\includegraphics{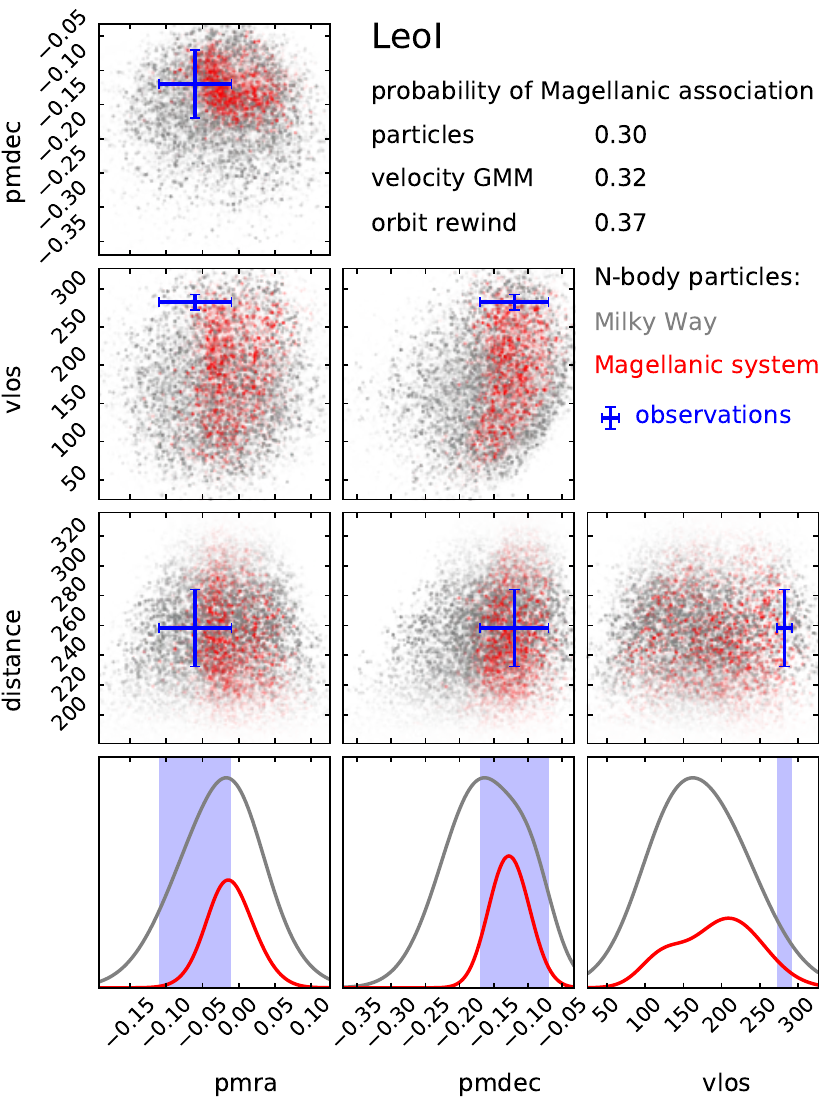}
\caption{
Examples of Gaussian mixture modelling of Milky Way (gray) and LMC (red) populations for two satellites, Fornax and Leo~I.
}  \label{fig:gmm}
\end{figure}

In the second method, we limit the matching to the spatial dimensions ($l$, $b$, $D$), obtaining a much larger sample (typically a few thousand particles) that are close to the target object in space. We select particles with sufficiently small Mahalanobis distance from the target object in the three spatial dimensions ($d_\mathrm{l,b,D}^{(i)}<4$), and their contribution to the distribution of each population $g$ (LMC or Milky Way) is weighted by the probability of spatial matching $p_\mathrm{l,b,D}^{(i)}\equiv \exp\big[-\frac{1}{2}(d_{l,b,D}^{(i)})^2 \big]$, so that the total mass of particles in each population is $M_g = \sum_{i\in g} m^{(i)}\,p_\mathrm{l,b,D}^{(i)}$. Then we construct two Gaussian mixture models, one for the Milky Way particles and the other for the particles originating in the Magellanic system. Each population is described by $N_\mathrm{comp}=2$ Gaussian components in the entire 6d space of observables, with amplitudes $A_{g,c}$, means $\overline{\boldsymbol X}_{g,c}$ and covariance matrices $\mathsf S_{g,c}$, where the sum of amplitudes for each population $g$ is $M_g$. Since particle weights are pre-multiplied by the probability of spatial matching, the distributions in these three dimensions are nearly Gaussian by construction; nevertheless, this procedure still allows one to recover correlations between spatial and kinematic dimensions, which are sometimes important for the LMC debris. The probability distribution of the given population $g$ (LMC or Milky Way) in the entire 6d space is given by 
\begin{eqnarray}
&\displaystyle p_g (\boldsymbol X) \propto \sum_{c=1}^{N_\mathrm{comp}}& A_{g,c}\;
\big|\det \mathsf S_{g,c} \big|^{-1/2} \times\\[-2mm]
&& \exp \Big[ -\frac{1}{2}(\boldsymbol X - \overline{\boldsymbol X}_{g,c})^T\, \mathsf S_{g,c}^{-1}\, (\boldsymbol X - \overline{\boldsymbol X}_{g,c}) \Big]. \nonumber
\end{eqnarray}
To account for observational uncertainties, this probability distribution should be convolved with the Gaussian error distribution described by the error covariance matrix $\mathsf{E}$, which for a Gaussian mixture model amounts to adding $\mathsf{E}$ to $\mathsf{S}_{g,c}$ in the above equation. Here the uncertainties in the spatial dimensions should be to a very large number (formally, infinity), since we have already preselected particles according to their spatial matching probability, so no further convolution is needed (this effectively amounts to marginalizing over these dimensions). Finally, the probability of the object to belong to $g$-th population is simply $p_g / \sum_g p_g$. The advantage of this method over the first one is that the smooth probability distributions are constructed from a much larger number of particles, but evaluated only in a possibly narrow range of kinematic coordinates, thus effectively interpolating them in a region where the actual number of matching particles could be rather small. The method is illustrated in Figure~\ref{fig:gmm} for two galaxies, Fornax and Leo~I.

The third method does not use the $N$-body snapshot directly, but instead relies on orbit integrations in the evolving gravitational potential described by a series of multipole expansions, as detailed in Section~\ref{sec:resimulation}. Namely, we sample $N_\mathrm{samp} = \mathcal O(10^3)$ points from the uncertainty distribution described by the mean coordinates $\boldsymbol X$ and the error covariance matrix $\mathsf{E}$. Then we integrate the orbits back in time from these initial conditions to the starting point of the simulation (10--11~Gyr ago), when the LMC was still approaching the Milky Way from a distance of a few hundred kpc. At this point, one may be tempted to simply count the fraction of orbits that end up being bound to the LMC at that time, as was done in most previous studies, but this will not give a correct answer, because not all points in our initial conditions are equally probable \textit{a posteriori}. To illustrate the fallacy, consider the case of very large PM uncertainties (e.g., Pisces~II) and/or missing line-of-sight velocity $V$ (e.g., Delve~2). Most of the rewound orbits would miss the LMC by a large margin, but they would also be unbound from the Milky Way or have apocentres well beyond the virial radius, making these orbits unlikely to appear in our simulation. Even if the uncertainties are small, the unweighted fraction of LMC-bound orbits tends to be underestimated, sometimes quite severely.
Instead, one needs to weigh each orbit with the probability of finding it among the initial conditions of our simulations, which consist of two equilibrium galaxy models. So we again use a mixture model approach, but instead of Gaussians in the present-day observable coordinates, we use the distribution functions of initial models $f_g(\boldsymbol x, \boldsymbol v)$ normalized such that the integral of $f_g$ over the entire 6d phase space is the total mass of each galaxy (or just its dark halo, in the case of Milky Way). Of course, in this evaluation we need to consider relative positions and velocities of particles with respect to each galaxy's centre $\{\boldsymbol x, \boldsymbol v\}_g^\mathrm{(cen)}$, and set $f_g$ to zero if the relative velocity exceeds the escape velocity. The un-normalized likelihood of association with either population is $p_g = \sum_{s=1}^{N_\mathrm{samp}} f_g\big(\boldsymbol x^{(s)} - \boldsymbol{x}_g^\mathrm{cen}, \boldsymbol v^{(s)} - \boldsymbol v_g^\mathrm{(cen)} \big)$, and the normalized probability of association is again $p_g / \sum_g p_g$.
This approach is similar to the one used in \citet{CorreaMagnus2022}, except that here we have a mixture of two populations bound to each galaxy, instead of just the Milky Way halo plus the ``unmixed'' population as in that paper. Since relatively few objects (mostly distant dSph) were found to be unmixed, and none of them are likely to be associated with the LMC, we ignore this component in the present study.

\begin{figure*}
\includegraphics{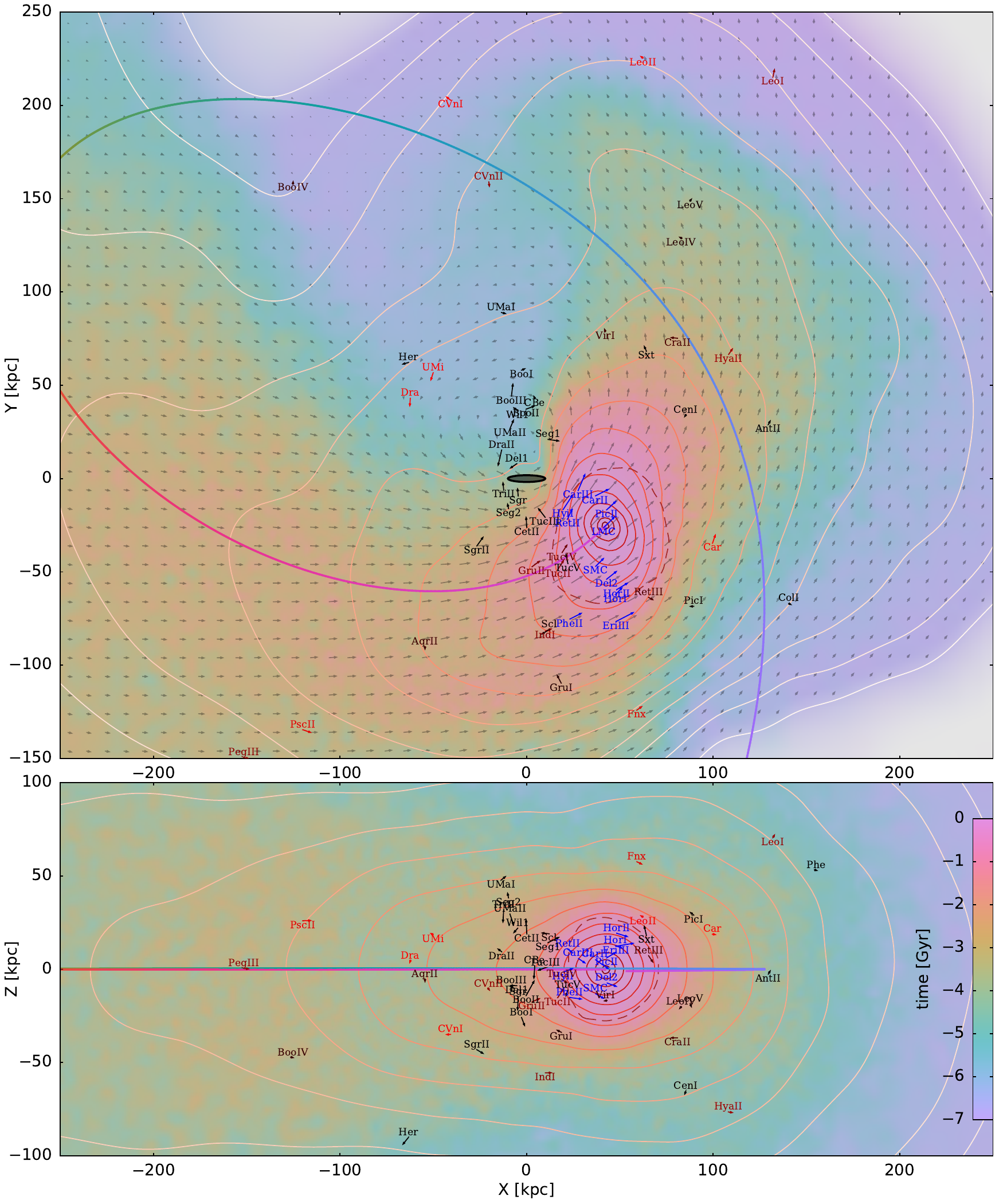}
\caption{
Spatial distribution and kinematics of LMC debris in the \L3--\M11 simulation. Top panel shows the view of the LMC orbital plane, which is almost perpendicular to the Galactic disc (whose projection onto this plane is shown by a gray ellipse), and bottom panel shows the edge-on projection of this plane. Density of LMC debris is shown by logarithmically spaced contours (four per decade), while the brown dashed contour roughly delineates the still bound region. Gray arrows show the mean velocity of particles, and black arrows with labels show the actual Milky Way satellites. Colour shading indicates the mean stripping time of debris, and the past orbit of the LMC is coloured in the same way.
}  \label{fig:debris}
\end{figure*}

\begin{figure}
\includegraphics{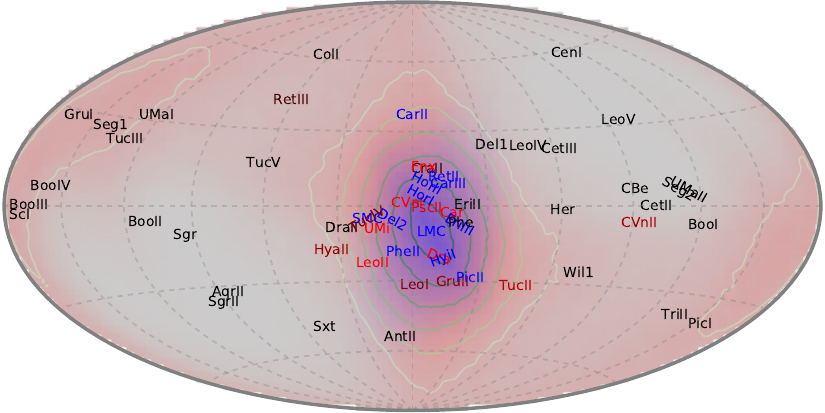}
\includegraphics{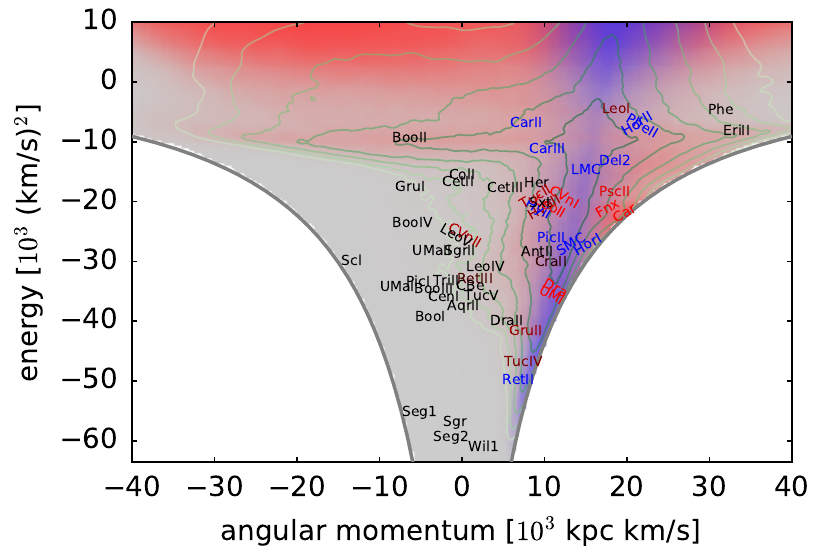}
\includegraphics{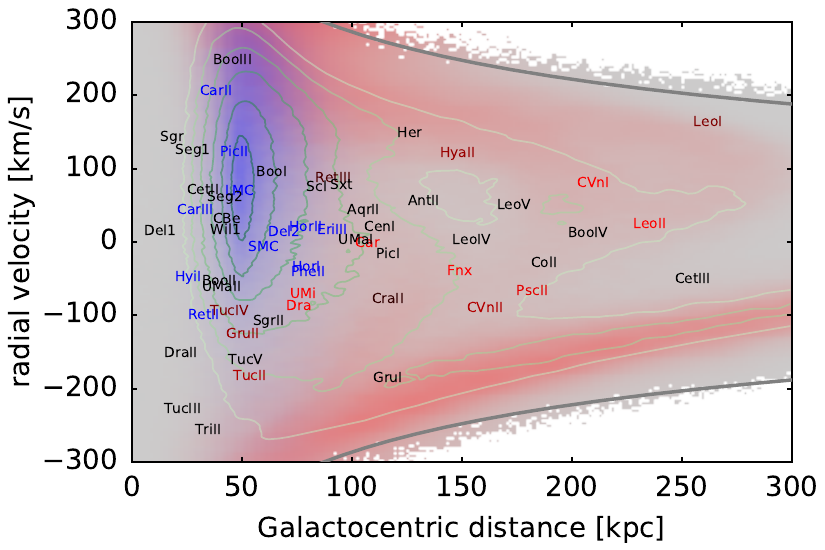}
\caption{
Distribution of LMC debris in the spaces of orbital poles (top panel), energy--angular momentum (middle panel) and distance--radial velocity (bottom panel) in the \L3--\M11 simulation. Orbital pole is the orientation of the particle's angular momentum in the Galactocentric coordinate system, where the Milky Way disc stars would be clumped around South pole (at the bottom of the celestial sphere), while the orbits of the LMC and many other satellites are roughly perpendicular to the disc plane. In the middle panel, energy is computed in the present-day Galactic potential, not taking into account the LMC debris itself, and the abscissa shows the angular momentum component parallel to that of the LMC. In all panels, contours show the distribution of LMC debris (logarithmically spaced with four contours per decade), and colour shading indicates whether the particles are bound to the LMC (blue), have been bound previously (red), or belong to the Milky Way halo (gray). Satellite galaxies are shown by names, for simplicity without displaying measurement uncertainties (see Figure~9 in \citealt{CorreaMagnus2022} for a variant of the orbital pole space with uncertainties for each satellite). LMC and its current satellites are displayed in blue, galaxies with significant probability of past association -- in shades of red (brighter for higher probability), other galaxies -- in black. We stress that our assessment of association probability is not based on these plots, but on the matching in the entire 6d of observable coordinates, and indeed some objects (e.g., Crater~II) appear unlikely to be associated even if they are located among the LMC debris in one or more panels above.
}  \label{fig:debris3}
\end{figure}

\begin{figure}
\includegraphics{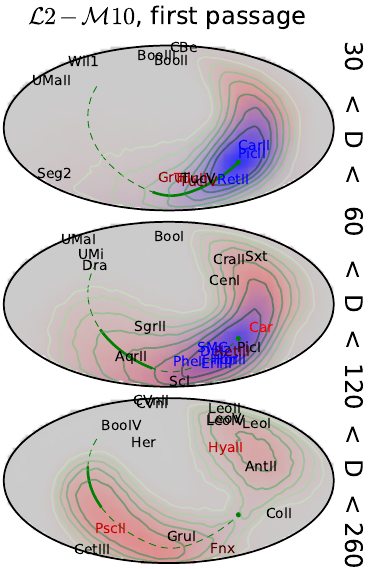}\hspace{9mm}
\includegraphics{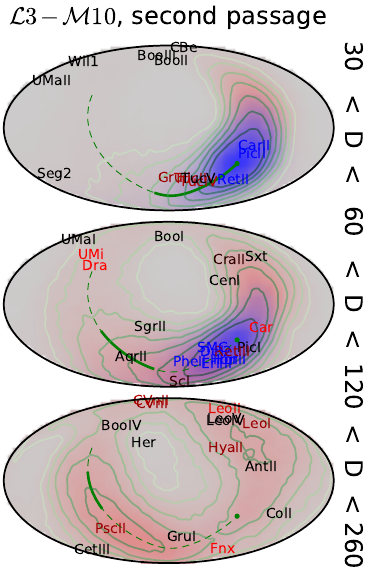}
\caption{
Distribution of LMC debris on the sky plane in three intervals of heliocentric distances, for the first passage (left column) and second passage (right column) simulations. Colour shading is the same as in Figure~\ref{fig:debris}, namely, particles still bound to the LMC are shown by blue, formerly bound -- by red, and Milky Way halo -- by gray; contours indicate the density of all particles originating in the Magellanic system. Satellites are shown in the respective distance intervals, coloured by probability of current (blue) / past (red) association with the Magellanic system. The differences between first and second-passage scenarios is not very dramatic overall, but most pronounced in the outermost distance bin around the north Galactic pole.
}  \label{fig:debris_skymap}
\end{figure}

\begin{figure*}
\includegraphics{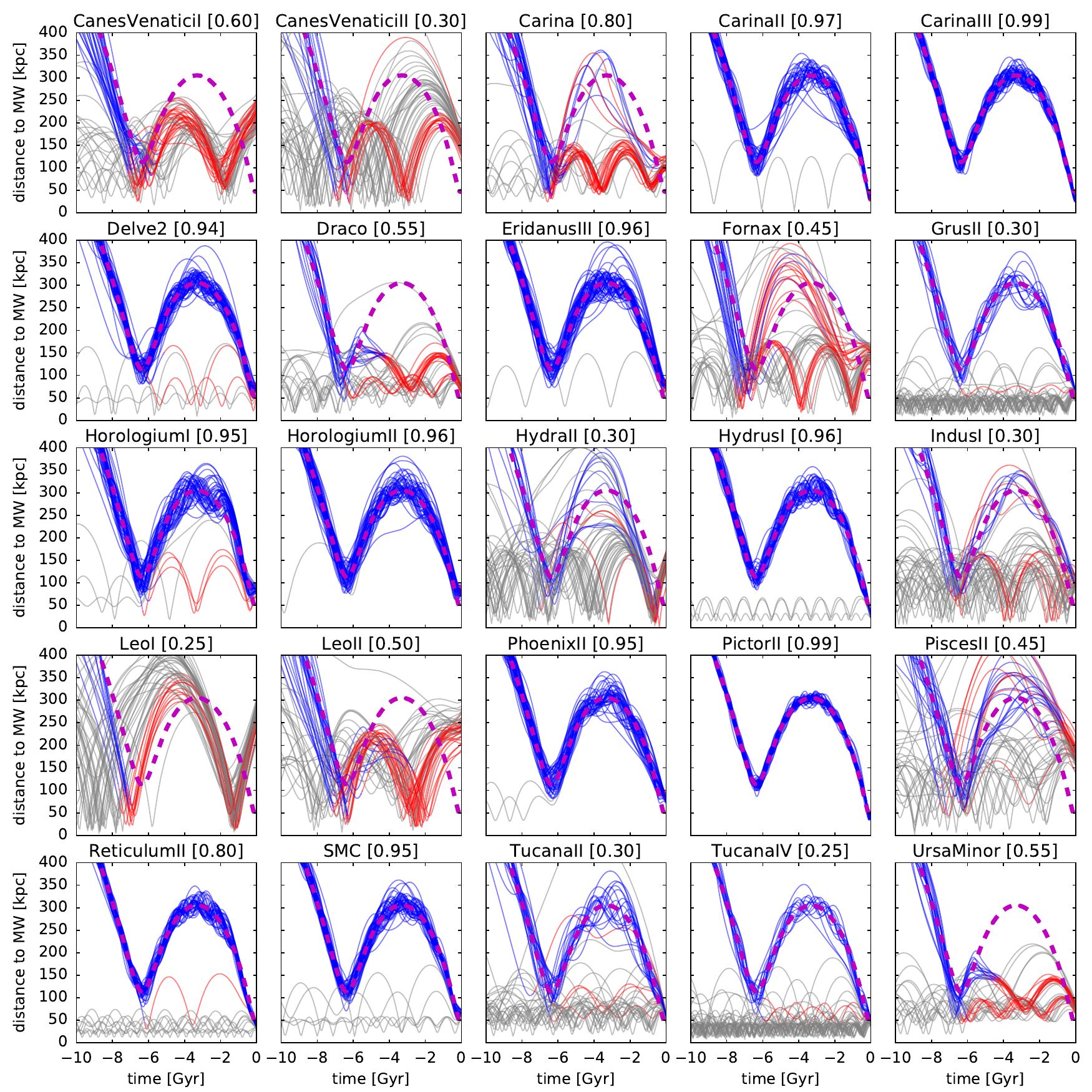}
\caption{
Past orbits of Milky Way satellites which have a significant probability of being associated with the LMC system. Shown are Galactocentric distances as a function of time for particles in the \L3--\M11 simulation that are compatible with the present-day position and velocity of each satellite, coloured according to its status: gray -- particles originating from the Milky Way halo, blue and red -- particles coming from the LMC (bound and unbound). Magenta dashed line shows the orbit of the LMC itself. The fraction of coloured orbits indicates the probability of Magellanic association, which is shown in brackets after each object's name.
}  \label{fig:satellite_orbits}
\end{figure*}

Reassuringly, all three methods agree remarkably well (to within 10\% difference in probability) for the vast majority of objects, so in the summary Table~\ref{tab:probability} we provide a single value (average between three methods) for each galaxy in each of the four simulations, skipping objects with less than 10\% probability of Magellanic association. These objects can be grouped into two classes. Eleven of them are likely to be Magellanic satellites at present time: Carina~II, Carina~III, Delve~2, Eridanus~III, Horologium~I, Horologium~II, Hydrus~I, Phoenix~II, Pictor~II, Reticulum~II and SMC. The probabilities exceed 90\% (except Reticulum~II, which is only $\sim$80\%), despite three of these galaxies lacking line-of-sight velocity measurements (Delve~2, Eridanus~III and Pictor~II). This classification broadly agrees with most recent studies (e.g., \citealt{Fritz2019}, \citealt{Erkal2020a}, \citealt{Patel2020}, \citealt{Battaglia2022}, \citealt{CorreaMagnus2022}, \citealt{Pace2022}, see Table~1 in \citetalias{Vasiliev2023} for a compilation). Figure~\ref{fig:debris} shows that these galaxies, shown by blue colour, are indeed located within the region occupied by particles still bound to the LMC (brown dashed contour) and move in the same direction.

Perhaps more interestingly, many of the remaining galaxies in Table~\ref{tab:probability}, shown by shades of red in the above figure, have a significant probability of association with the LMC, even though they are not currently bound to it. In particular, four of the classical dSph -- Carina, Draco, Fornax and Ursa Minor -- have a roughly equal chance to either be accreted with the Magellanic system or originate from elsewhere (in case of Carina, the chances are even higher, $\sim 2:1$). A distant galaxy Leo~II has a probability of Magellanic association between 0.3 and 0.5, and even the peculiar Leo~I, the least bound object in this sample, has a non-negligible chance of association in the \L3 series of models\footnote{We remind that regardless of whether Leo~I is associated with the LMC or not, its current velocity relative to the Milky Way centre has been increased by a few tens \kms due to the reflex motion of the Milky Way caused by the recent LMC flyby \citep{Erkal2020b, CorreaMagnus2022}.}. Even some galaxies with missing velocity components might be associated with the LMC under certain conditions; Table~\ref{tab:predicted} lists the predicted PM and line-of-sight velocity values assuming such an association.
Summing up the probabilities of galaxies in this group, we may conjecture that the original Magellanic system might have brought in some 4--6 galaxies that became Milky Way satellites, half of them in the mass range of classical dSph. They would populate the unusual 12 magnitudes luminosity gap between the SMC and other ultrafaint LMC satellites highlighted by \citet{Dooley2017} and \citet{Pardy2020}, and indeed the latter study suggested that Carina and/or Fornax might fill this gap.

Figure~\ref{fig:debris3} shows the distribution of Magellanic and Milky Way particles in different kinematic spaces: orbital pole orientation (top panel), energy--angular momentum (middle panel) and Galactocentric radius--velocity (bottom panel), and the locations of observed galaxies in these spaces. Particles still bound to the LMC and galaxies from the first group are well localized in each of these spaces, but possible former LMC satellites are considerably scattered and mixed with native Milky Way satellites in any particular projection, justifying the need for a full 6d analysis as in the present study. The distribution of former satellites in orbital phase is quite asymmetric, with all but one object (Pisces~II) populating the leading arm of the tidal debris stream (the trailing arm is visible as the red band in the lower part of the bottom panel). A similar situation occurs with the globular clusters stripped from the Sagittarius galaxy, and the reason for this asymmetry is currently unknown. As the census of Milky Way satellites is very likely still incomplete \citep{Koposov2009,Jethwa2016}, one may use these ``finding charts'' to assess the possibility of Magellanic association of any object discovered in the future, though only if we have kinematic information. Figure~\ref{fig:debris_skymap} instead shows just the observable sky-plane distribution of LMC debris in different distance bins, both for first-passage (left) and second-passage (right) simulations. Unfortunately, the difference between these scenarios is not very dramatic, and mostly seen in the outermost distance bin and in the Northern hemisphere, which cannot be reached by debris from a single passage.

Figure~\ref{fig:satellite_orbits} shows possible past orbits of selected galaxies in one of the simulations (\L3--\M11), illustrating that the galaxies from the second group have been stripped after the first pericentre passage, and their subsequent orbits do not resemble that of the LMC, but still stay roughly in the same orbital plane. This kinematic peculiarity, of course, has been known for a long time, as discussed in Section~\ref{sec:satellites_literature}, but here we provide an explicit demonstration that such an arrangement is a natural outcome of the tidal stripping of a massive Magellanic system.

\begin{table}
\caption{Galaxies with a significant probability of association with the LMC system in the four simulations with different LMC masses and Milky Way potentials. Current satellites of the LMC are shown in italics. We also list the absolute magnitudes and heliocentric distances from the catalogue of \citet[updated in 2021]{McConnachie2012}.
}  \label{tab:probability}
\hspace{-5mm}\begin{tabular}{lrrllll}
Name            &$M_V$  & $D$ &\makebox[7mm][l]{\!\!\!\!\L2\M10}&\makebox[7mm][l]{\!\!\!\!\L2\M11}&\makebox[7mm][l]{\!\!\!\!\L3\M10}&\makebox[7mm][l]{\!\!\!\!\L3\M11}\\[1mm]
Canes Venatici  I\!\!\!&$-8.6$ &\!210 & 0.1   & 0.3   & 0.3   & 0.6  \\
Canes Venatici II\!\!\!&$-4.6$ &\!160 & 0.3   & 0.35  & 0.35  & 0.3  \\
Carina                 &$-8.6$ &\!106 & 0.55  & 0.65  & 0.7   & 0.8  \\
{\it Carina II }       &$-4.5$ &  37  & 0.97  & 0.95  & 0.98  & 0.97 \\
{\it Carina III}       &$-2.4$ &  28  & 0.99  & 0.99  & 0.99  & 0.99 \\
Crater II              &$-8.2$ &  117 & 0.2   & 0.1   & 0.1   & 0.1  \\
{\it Delve 2}          &$-2.1$ &  71  & 0.93  & 0.91  & 0.95  & 0.94 \\
Draco                  &$-8.7$ &  76  & 0.4   & 0.45  & 0.55  & 0.55 \\
{\it Eridanus III}     &$-2.3$ &  91  & 0.95  & 0.95  & 0.97  & 0.96 \\
Fornax              &\!$-13.4$ &\!147 & 0.4   & 0.4   & 0.5   & 0.45 \\
Grus II                &$-3.9$ &  55  & 0.25  & 0.25  & 0.3   & 0.3  \\
{\it Horologium I}     &$-3.5$ &  79  & 0.92  & 0.90  & 0.94  & 0.95 \\
{\it Horologium II}    &$-1.5$ &  78  & 0.95  & 0.94  & 0.97  & 0.96 \\
Hydra II               &$-4.8$ &\!151 & 0.2   & 0.3   & 0.3   & 0.3  \\
{\it Hydrus I}         &$-4.7$ &  28  & 0.95  & 0.95  & 0.96  & 0.96 \\
Indus I                &$-1.5$ &\!105 & 0.2   & 0.2   & 0.3   & 0.3  \\
Leo I               &\!$-12.0$ &\!258 & 0.15  & 0.1   & 0.3   & 0.25 \\
Leo II                 &$-9.6$ &\!233 & 0.3   & 0.4   & 0.4   & 0.5  \\
{\it Phoenix II}       &$-3.3$ &  83  & 0.95  & 0.93  & 0.96  & 0.95 \\
{\it Pictor II}        &$-4.2$ &  46  & 0.99  & 0.99  & 0.99  & 0.99 \\
Pisces II              &$-4.1$ &\!183 & 0.2   & 0.3   & 0.35  & 0.45 \\
{\it Reticulum II}     &$-3.6$ &  31  & 0.8   & 0.75  & 0.85  & 0.8  \\
Reticulum III          &$-3.3$ &  92  & 0.1   & 0.1   & 0.25  & 0.15 \\
{\it SMC}           &\!$-16.8$ &  63  & 0.92  & 0.92  & 0.95  & 0.95 \\
Tucana II              &$-3.9$ &  58  & 0.3   & 0.2   & 0.35  & 0.3  \\
Tucana IV              &$-3.5$ &  47  & 0.2   & 0.2   & 0.3   & 0.25 \\
Ursa Minor             &$-8.4$ &  76  & 0.45  & 0.45  & 0.45  & 0.55 \\
Virgo I                &$-0.8$ &  91  & 0.1   & 0.1   & 0.1   & 0.1  \\
\end{tabular}
\end{table}

\begin{table}
\caption{Predicted values of PM $\mu_\alpha$, $\mu_\delta$ (\masyr) and/or line-of-sight velocity (\kms) for objects with missing or highly uncertain measurements, assuming their Magellanic association; measured values are replaced by a box.
}  \label{tab:predicted}
\begin{tabular}{lccc}
Name & $\mu_\alpha$ & $\mu_\delta$ & $v_\mathrm{los}$ \\[1mm]
Delve~2     & $\Box$          & $\Box$          & $ 150\pm 80$ \\
Eridanus~III& $\Box$          & $\Box$          & $ 160\pm 70$ \\
Indus~I     & $ 0.0 \pm 0.2$  & $-1.0 \pm 0.2$  & $ -20\pm 80$ \\
Pegasus~III & $ 0.12\pm 0.06$ & $-0.27\pm 0.06$ & $\Box$       \\
Pictor~I    & $\Box$          & $\Box$          & $ 200\pm 80$ \\
Pictor~II   & $\Box$          & $\Box$          & $ 320\pm 90$ \\
Pisces~II   & $ 0.2 \pm 0.07$ & $-0.33\pm 0.07$ & $\Box$       \\
Virgo~I     & $-0.2 \pm 0.1$  & $-0.2 \pm 0.1$  & $ 300\pm 40$ \\
\end{tabular}
\end{table}

\begin{figure}
\includegraphics{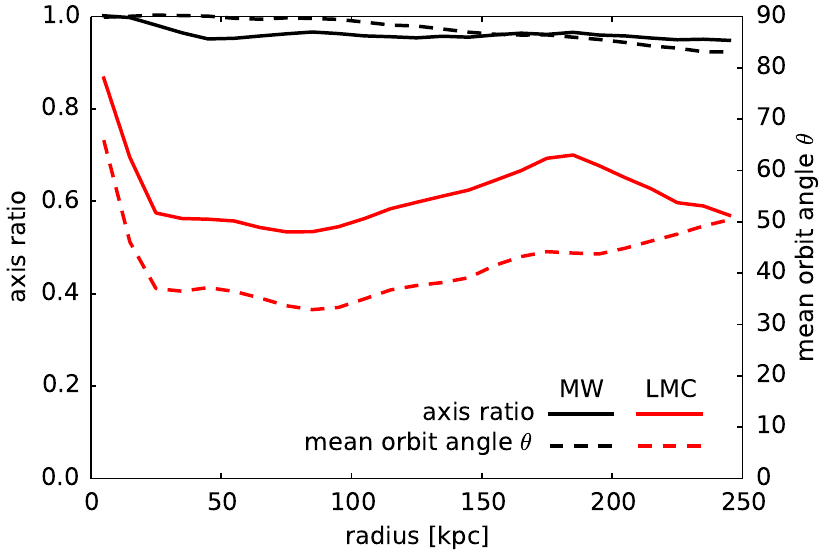}
\caption{
Spatial (solid) and kinematic (dashed) asymmetries in the distribution of Magellanic (red) and Milky Way halo particles (black) in the \L3--\M10 simulation. Let $\boldsymbol n$ be the unit vector in the direction of the LMC angular momentum. For a particle located at position $\boldsymbol r$, the distance from the LMC orbital plane is $z'\equiv \boldsymbol r \cdot \boldsymbol n$, and the axis ratio of the moment of inertia tensor, shown by solid lines, is $\Big({\langle z'^2 \rangle} \Big/ {\langle r^2-z'^2 \rangle}\Big)^{1/2}$, where the angle brackets denote averaging over an ensemble of particles. Likewise, the angle between the particle's angular momentum $\boldsymbol L$ and that of the LMC is $\theta \equiv \arccos(\boldsymbol L \cdot \boldsymbol n)$, and the mean value of this angle is shown by dashed lines.
For the particles originating in the Milky Way halo, both quantities deviate only slightly from the initial values for a spherically symmetric isotropic halo, namely axis ratio of unity and mean angle of $90^\circ$. By contrast, Magellanic debris (including particles still bound to the LMC) are significantly concentrated towards the orbital plane (axis ratio of 0.5--0.6 and orbital angle of 30--40$^\circ$).
}  \label{fig:vpos}
\end{figure}

\begin{figure}
\includegraphics{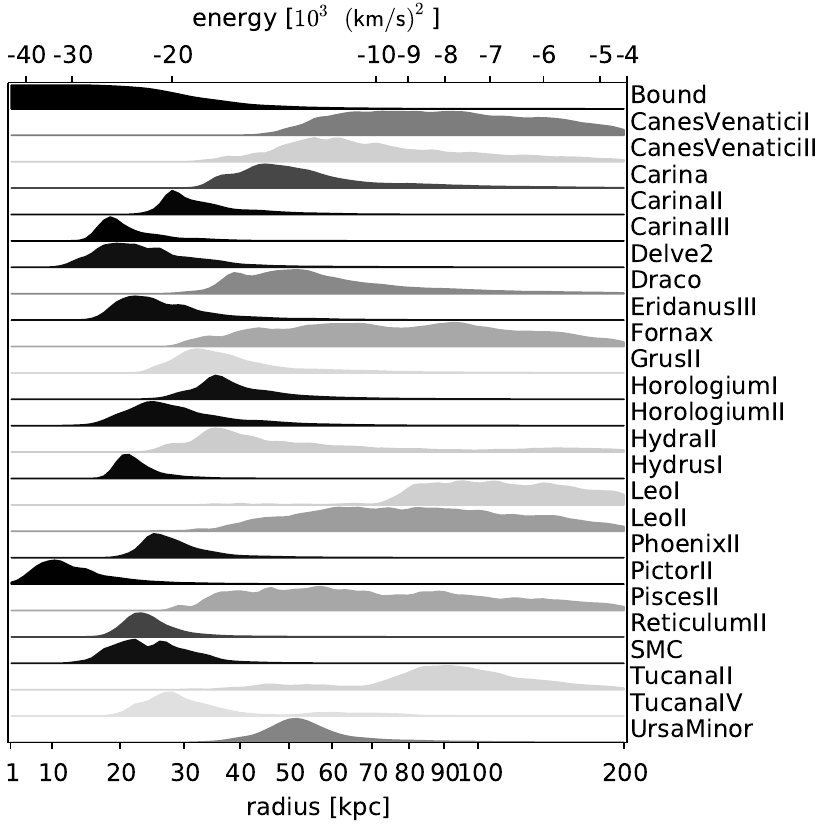}
\caption{
Possible initial locations of LMC satellites for the \L3--\M11 simulation. Shown are the probability distributions of initial conditions for objects with high chance of LMC association, obtained by orbit rewinding with posterior reweighting (third method described in Section~\ref{sec:satellite_membership}). For each object, we sample present-day positions and velocities from the observational uncertainties and rewind the orbits back to the initial time of the simulation in the pre-recorded time-dependent potential. Each orbit that was initially bound to the LMC is assigned a value $\varepsilon\in[0,1]$, which is the percentile of its initial energy in the energy distribution of LMC-bound particles. The histograms of $\varepsilon$ are shown for each object in a separate row, shaded by the overall association probability (darker is higher). Top row shows the histogram of all particles currently bound to the LMC; if we were to plot the histogram of all LMC particles, it would be flat by construction. For convenience, the values of $\varepsilon$ are translated to the initial energy (top axis) and initial radius of a circular orbit with this energy (bottom axis), both quantities having a non-linear dependence on $\varepsilon$. Unsurprisingly, current LMC satellites have to originate from more bound orbits (within $\sim 40$~kpc) in order to avoid being stripped on the first passage.
}  \label{fig:initenergy}
\end{figure}

\citet{Pawlowski2011} proposed that the disc of satellites consists of tidal dwarf galaxies formed out of gas that was tidally stripped from the progenitor galaxy (possibly the proto-Magellanic system) during its pericentre passage. In our view, this scenario is conceptually similar to the one considered in the present paper, except that we assume that these galaxies were already formed before the first pericentre passage of the Magellanic system, but the stripping mechanism is the same, so we remain agnostic as to whether these galaxies are expected to be ``primordial'' (formed in the standard cosmological paradigm) or ``second generation'' (formed out of tidally stripped gas without necessarily be embedded in dark matter haloes). Full hydrodynamical simulations may be needed to resolve this question.

Although thin satellite planes appear to be quite rare around generic Milky Way-mass host galaxies in cosmological simulations (see Section~2.2.3 in \citealt{Pawlowski2021} for a review of relevant literature), \citet{Samuel2021} found that the fraction of satellite planes significantly increases if one selects host galaxies that have an LMC analogue near its pericentre. \citet{GaravitoCamargo2021b} suggested that the dynamical perturbation from the LMC may enhance the apparent clustering of orbital poles due to direct deflection of orbits by the LMC flyby and the global displacement of the Milky Way-centred reference frame relative to the outer halo. On the other hand, \citet{Pawlowski2022} and \citet{CorreaMagnus2022} found that this effect alone is insufficient to account for the observed VPOS thinness. After excluding the current LMC satellites and rewinding the orbits of remaining objects in the combined Milky Way--LMC potential back in time to undo the LMC perturbation, the clustering of orbital poles was not significantly affected. 
To complement this analysis, Figure~\ref{fig:vpos} shows the asymmetries in the spatial distribution (axis ratio of the moment of inertia tensor) and in kinematics (mean angle $\theta$ between the angular momentum of particles relative to that of the LMC) of particles in our simulations, separately for the Milky Way (black) and Magellanic particles (red). It is clear that the perturbations in the native Milky Way halo population are minimal -- the orbital poles of particles stay very close to isotropic ($\theta \approx 90^\circ$) within 100 kpc and only show a slight enhancement in the direction of the LMC orbital pole further out, confirming the results of \citet{Pawlowski2022}. By contrast, particles belonging to the Magellanic debris have a prominent concentration of their orbital poles ($\theta\lesssim 40^\circ$) and are more flattened (axis ratio 1:2), although the observed satellite plane is still thinner.

However, these idealised simulations may not be representative of the evolution of Galactic satellite distribution in a realistic cosmological context. Following upon the work of \citet{Samuel2021}, \citet{GaravitoCamargo2023} considered Milky Way--LMC analogues in the \textit{Latte} cosmological simulations, and found that the enhancement of orbital pole concentration is more prominent than in isolated simulations of Milky Way--LMC encounter, although the displacement of the inner Galaxy w.r.t.\ the outer halo remains the driving mechanism. However, this enhancement persist only for a short time around the pericentre passage of the LMC analogue: using the \textit{IllustrisTNG} cosmological simulations, \citet{Kanehisa2023} found that such encounters do not produce long-lived satellite planes. This result is at odds with our conclusion that the objects stripped on the first passage remain close to the LMC orbital plane; the discrepancy might be caused by the idealised (non-cosmological) nature of our simulations.

Figure~\ref{fig:initenergy} shows the possible origins of galaxies that were part of the Magellanic system, expressed as the percentile of their initial binding energy among all LMC particles. Naturally, the eleven galaxies from the first group (current LMC satellites) must have started rather deep inside its potential in order to avoid being stripped on the first passage, but not too tightly bound (with the exception of Pictor~II) to avoid tidal disruption by the LMC itself.  Galaxies from the second group have generally broader distributions of possible initial binding energies, but still avoiding least bound orbits, which may explain their stronger spatial and dynamical coherence compared to the entire distribution of Magellanic debris shown on the previous figure.

We also examined the probability of Magellanic association of dSphs for the simulation of a single passage (\L2\M10 run for 4~Gyr). In this case, all eleven galaxies from the first group are still highly likely to be current LMC satellites, but among the second group, only Carina retains a significant (40--50\%) probability of association, Grus~II, Hydra~II, Pisces~II and Tucana~II stay at the level 20--40\%, and remaining galaxies are not compatible with the distribution of Magellanic debris. Thus the ability to explain the existence of the satellite plane is an important argument in favour of the second-approach model in the standard cosmological paradigm. The alternative scenario advocated by \citet{Pawlowski2013} and subsequent studies involves modified Newtonian gravity theories (e.g., MOND, \citealt{Milgrom1983}), in which case it is the early encounter with M31 rather than LMC that could give rise to the dynamically coherent second-generation tidal dwarf galaxy population \citep{Bilek2018,Banik2022}.

%%%%%%%%
\section{Discussion}  \label{sec:discussion}

%We propose a second-approach scenario for the Magellanic system.
Although most studies in the last decade concentrated on the first passage (or recent infall) scenario for the Magellanic clouds, in this study we demonstrate that the alternative scenario in which they had another pericentre passage 5--10 Gyr ago does not violate any observational constraints:
\begin{itemize}
\item The updated measurement of the LMC PM with HST \citep{Kallivayalil2013} and \Gaia \citep{Luri2021} led to a downward revision of the LMC tangential velocity by a few tens \kms compared to the earlier HST measurement by \citet{Kallivayalil2006a}. Whereas the latter would make the LMC orbit unbound unless the Milky Way mass is well above $1.5\times10^{12}\,M_\odot$, the updated PM makes it comfortably bound even if $M_\mathrm{MW}$ is around $10^{12}\,M_\odot$, as preferred by most recent estimates (Section~\ref{sec:orbit}).
\item A high initial LMC mass ($M_\mathrm{LMC} \gtrsim 2\times10^{11}\,M_\odot$) shortens the inferred orbital period, but increases the distance of the previous pericentre passage to $\gtrsim 100$~kpc. At this distance, the tidal radius of the LMC is large enough that it has no problem of retaining all satellites that are currently still orbiting it, including the SMC (Section~\ref{sec:satellite_membership}).
\item The perturbations of the Milky Way halo induced by the LMC during its current flyby look nearly the same in the first- and second-passage scenarios, and qualitatively agree with the observed signatures, though some deviations remain unexplained (Section~\ref{sec:mwhalo}).
\end{itemize}

The latter circumstance is rather unfortunate, as one would like to have a definite prediction or observed signature that would distinguish the two scenarios. Although we do not have a smoking-gun evidence in favour of the second-passage scenario, a strong argument in its support is the natural emergence of the spatially and kinematically coherent plane of satellites as a product of tidal stripping during the previous pericentre passage. 
We stress that our analysis does not aim to quantify how common or unusual is this structure -- it may well be that the Milky Way--LMC system is rather special \citep[e.g.,][]{Busha2011}. It also does not address the likelihood of a particular configuration of objects within the orbital plane of the LMC: for instance, the results would not change if all galaxies were concentrated in the same spatial region rather than being uniformly spread in angles. It merely evaluates the conditional probability of a given object to come from the Magellanic system, given its current position and velocity and a plausible history of interaction between the Milky Way and the LMC. We find that in the second-passage scenario, most galaxies with orbital poles similar to that of the LMC indeed have a large probability of past association, whereas if the LMC is on its first approach, only Carina dSph could be possibly coming alongside it.

Our simulations and analysis have a number of limitations and simplifying assumptions, some of which can be addressed in the follow-up work:
\begin{itemize}
\item We only considered a small number of models, focusing on the dependence of the orbital period on the Milky Way and LMC masses. It would be interesting to vary the shape of the Galactic potential, as it also has a significant effect on the orbital period (Figure~4 in \citetalias{Vasiliev2023}).
\item We modelled the LMC as a single spherical halo, without explicitly introducing a stellar component. The actual LMC has a (thick) stellar disc with a bar, whose formation might have been triggered by the previous pericentre passage \citep[e.g.,][]{Lokas2014}, although it is more likely to result from its interaction with the SMC. 
\item Speaking of the latter, the neglect of the gravitational pull from LMC's most massive satellite does have a substantial effect on its inferred orbit. Although some studies (e.g., \citealt{Jethwa2016}, \citealt{Patel2020}) considered the three-body interaction of Milky Way, LMC and SMC using simple orbit integrations, it has not yet been studied in full $N$-body simulations. The problem of matching the current position and velocity of both Magellanic Clouds could be too challenging, so a simulation of the live SMC in the pre-recorded time-dependent potential of the Milky Way and LMC (similar to the simulations of Sagittarius dSph conducted in \citealt{Vasiliev2021}) might be an intermediate step. Even if the SMC is much less massive than the LMC, it still shifts the centre of mass of both Clouds and reduces their Galactocentric orbital period (Figure~3 in \citetalias{Vasiliev2023}).
\item Our simulations were pure $N$-body, but the Clouds contain a substantial amount of gas. In the currently dominant first-passage scenario, the Magellanic gas stream results from the interaction of the two Clouds \citep{Besla2010}. It would be very interesting to examine the formation mechanisms of the Magellanic stream in the second-passage scenario.
\item It is also rather inconvenient that one has to run an entire $N$-body simulation to reliably reconstruct the past trajectory of the LMC, as the approximate orbit integration method, while very fast, is not accurate enough (Figure~\ref{fig:orbit_approx}). A yet-to-be-developed hybrid method for approximating the orbital and mass loss evolution that sits between these two extremes would be very valuable for extensive exploration of parameter space (e.g., of the Milky Way potential).
\item The analysis of observational constraints on the star formation histories of the Magellanic Clouds and other dwarf galaxies may help to constrain their past orbital histories \citep[e.g.,][]{Hasselquist2021,Mazzi2021,Massana2022}, supporting or disfavouring the proposed scenario.
\item The reconstruction of past trajectories of satellites assumed them to be point masses, i.e., ignored the tidal stripping. Some of the LMC satellites might not have survived on their inferred orbits with fairly small pericentre distances and short periods, though the range of possible past orbits is usually large enough to accommodate less stressful ones. Still, the evolution of LMC satellites in the tidal field is worth exploring in more detail.
\end{itemize}

To facilitate future studies of these and other related questions, we provide the simulation snapshots, time-dependent potentials, scripts for their analysis and for tailoring the orbital initial conditions, in a public Zenodo repository. For instance, one could use them for quickly evaluating the probability of Magellanic association for any newly discovered satellites (with or without kinematic information).

It is clear that the presence of the Magellanic Clouds in such a special configuration (near their pericentre) is a rather peculiar feature of Milky Way dynamics. As demonstrated in this study, even the basic question about when they have been accreted does not have a definite answer at the current level of knowledge, though we provided arguments in favour of the second-passage scenario. That the answer is highly sensitive to various parameters makes them an excellent tool for exploring the structure and past history of our Galaxy, and the Clouds are very welcome even for that reason alone, not to mention their sheer beauty.

\textbf{Acknowledgements:} I thank many of my colleagues at Edinburgh and Cambridge for stimulating discussions and comments, in particular, Mike Petersen, whose request to share the script for fitting the orbital initial conditions prompted me to significantly improve it, in turn enabling a reliable analysis of second-passage orbits. I also thank the referee for valuable comments that helped to improve the presentation.

\textbf{Data availability:} The simulation data and associated scripts are provided at \url{https://zenodo.org/record/8015660}.

%%%%%%

%%%%%%%%%
\appendix
\section{Procedure for finding the orbital initial conditions}  \label{sec:appendix}

In this section, we detail the solutions to technical problems appearing in the process of finding the orbital initial conditions that deliver the LMC to the correct position with correct velocity at the end of simulation. There are three important aspects of the procedure: determination of smooth trajectories of both galaxies, nonlinear transformation of phase-space coordinates, and computation of the Jacobian for the Newton method. The present procedure is an improved version of the one described in the appendix of \citet{Vasiliev2020}.

\subsection{Extracting the smooth trajectories of Milky Way and LMC from the simulation}

The initial data for orbit determination comes from recording position $\tilde{\boldsymbol x}_\mathrm{MW/LMC}$ and velocity $\tilde{\boldsymbol v}_\mathrm{MW/LMC}$ of each galaxy's centre at every block timestep of the simulation (8~Myr). These quantities are defined as median values among particles within a radius $r_\mathrm{max}=10$~kpc from the centre of the corresponding galaxy. Due to this self-referencing definition, we compute these quantities iteratively, starting with the linearly extrapolated centre position from the previous timestep, computing the median coordinates of particles within $r_\mathrm{max}$, setting this as an updated centre, and repeating the procedure a few times. However, even in the highest-resolution simulations, these measurements are noisy and must be smoothed before being used for computing the Jacobian and the acceleration of the Milky Way-centered reference frame. Our initial attempts at using spline smoothing were not very fruitful, at least in low-resolution simulations ($N_\mathrm{body}=0.5\times10^6$): it was impossible to find a good balance between reducing the noise and still recovering relevant rapid variations in the trajectories near pericentre passages. Instead, an alternative method was designed that took advantage of the particular physical setup.
It starts from a somewhat inconsistent but often used approximation that the two galaxies are rigid structures with potentials $\Phi_\mathrm{MW}(\boldsymbol x)$, $\Phi_\mathrm{LMC}(\boldsymbol x)$, but at the same time their centres satisfy the equations of motion for a test particle in the other galaxy's potential:
\begin{equation*}
\begin{array}{ll}
\dot{\boldsymbol x}_\mathrm{LMC} = \boldsymbol v_\mathrm{LMC},\quad &
\dot{\boldsymbol v}_\mathrm{LMC} = -\nabla\Phi_\mathrm{MW}(\boldsymbol x_\mathrm{LMC} - \boldsymbol x_\mathrm{MW}) + \boldsymbol a_\mathrm{res,\,LMC}, \\[1mm]
\dot{\boldsymbol x}_\mathrm{MW} = \boldsymbol v_\mathrm{MW},\quad &
\dot{\boldsymbol v}_\mathrm{MW} = -\nabla\Phi_\mathrm{LMC}(\boldsymbol x_\mathrm{MW} - \boldsymbol x_\mathrm{LMC}) + \boldsymbol a_\mathrm{res,\,MW}. \\[1mm]
\end{array}
\end{equation*}
The extra terms in the right-hand side $\boldsymbol a_\mathrm{res}$ are the residual accelerations, introduced as placeholders for all factors not accounted for in this simple model, such as dynamical friction and self-force from the deformed LMC. They have a significantly lower amplitude and temporal frequency than the primary accelerations, which are computed from the analytic initial potentials of both galaxies. These residual accelerations are represented by cubic B-splines with strategically placed grid nodes (more closely spaced near orbit pericentres) and free coefficients varied during the fit. Namely, the measured (tilded) positions and velocities are fit in the least-square sense by the model that results from the integration of the above equations of motion, including the corrections introduced by the residual accelerations. The fit is mildly nonlinear, because the primary accelerations depend on the unknown smooth trajectories of both galaxies, but a good first approximation is obtained by putting the measured positions as the argument of potential gradients, and then iterating one more time with the smoothed positions. This procedure produces smooth and physically realistic trajectories with only $\mathcal O(10)$ free parameters of the spline fits to residual accelerations for each galaxy.

We note that the above equations for both galaxy trajectories can be combined into an equation for the relative trajectory $\Delta\boldsymbol x \equiv \boldsymbol x_\mathrm{LMC} - \boldsymbol x_\mathrm{MW}$, $\Delta\boldsymbol v \equiv \boldsymbol v_\mathrm{LMC} - \boldsymbol v_\mathrm{MW}$ of a test particle moving in the combined gravitational potential of both galaxies \textit{centred at origin} $\Phi(\Delta\boldsymbol x) \equiv \Phi_\mathrm{MW}(\Delta\boldsymbol x) + \Phi_\mathrm{LMC}(-\Delta\boldsymbol x)$, just as in a regular two-body problem. The only reason for treating them separately at this stage is to determine the acceleration of the Milky Way-centred reference frame as the second derivative of its trajectory. The next two stages operate with the relative coordinates $\{\Delta\boldsymbol x, \Delta\boldsymbol v\}$ only.

\subsection{Nonlinear coordinate transformation}

Despite the great increase in smoothness of orbits delivered by the above procedure, there is still a limit on the achievable precision of phase-space coordinates of the LMC. Simply put, we need the companion orbits to be sufficiently different from the master orbit to have reliable measurement of their displacement. The problem is that orbits started with an initial offset of order 1~kpc or 1~\kms accumulate a significant phase difference after 10~Gyr of evolution, so that the present-day offsets could easily measure in tens of kpc, making the Jacobian extremely ill-conditioned. On the other hand, the companion orbits still remain close to the master orbit and pass close to the target point $\boldsymbol w_\mathrm{true}$ (the present-day position and velocity of the LMC), just at different times. 

To address this problem, we introduce a two-step transformation of phase-space coordinates. 
In the first step, we define a new basis in the 6d phase space associated with a given point $\big\{\boldsymbol x^\mathrm{(b)}, \boldsymbol v^\mathrm{(b)} \big\}$ on the trajectory (initial or final). To implement rotations in the 6d phase space, we need to express both position and velocity in the same units, which is achieved by introducing a scaling factor $\tau$ with a dimension of time; it should be of order the dynamical time for a given orbit, e.g., 1~Gyr. 
The first basis vector points along the direction of travel:
$\boldsymbol b^{(1)} \propto \big\{\boldsymbol v^\mathrm{(b)},\; -\tau\,\nabla\Phi(\boldsymbol x^\mathrm{(b)}) \big\}$, where $\Phi$ is the combined potential of both galaxies. The second basis vector points along the gradient of the Hamiltonian: $\boldsymbol b^{(2)} \propto \big\{ \tau\,\nabla\Phi(\boldsymbol x^\mathrm{(b)}),\; \boldsymbol v^\mathrm{(b)} \big\}$. Obviously, these two basis vectors are orthogonal by construction, and are normalized to unit length. A shift along the first basis vector keeps the particle on the same trajectory (in as much as the approximate equations of motion are valid), but at an earlier or a later moment of time, i.e., changes the orbital phase. On the other hand, a shift along the second basis vector puts it onto a different orbit with higher or lower energy, but approximately keeping the orbital phase. We then use the same pattern to define the remaining four basis vectors: two of them correspond to changes in the other integrals of motion (total angular momentum and the orientation of the orbital plane), and two other vectors represent shifts in the corresponding phase angles. Displacements along each of the basis vectors except the first one correspond to different orbits, so we may label an orbit by the coordinates of the point at which it crosses the ``fiducial hyperplane'' formed by these five vectors.

\begin{figure}
\includegraphics{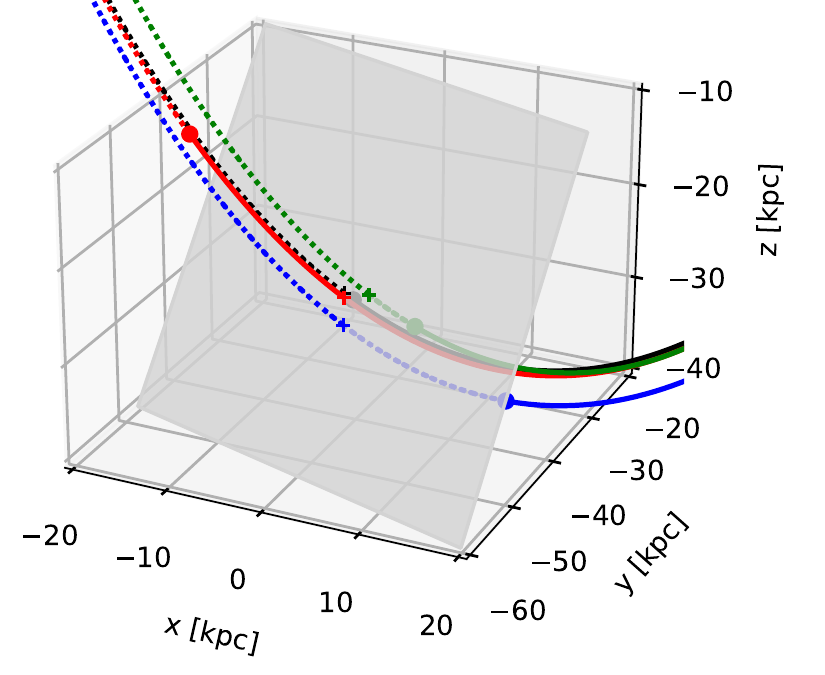}
\caption{
Simplified illustration of the nonlinear coordinate transformation. Shown are the master orbit in black and three companion orbits in different colours, and past/future orbit segments rendered by solid and dotted lines. Dots mark the present-day positions of the LMC in these simulations, which differ by a few tens kpc, but most of this difference stems from the time offset, as the trajectories themselves are quite close. To make use of this proximity, we compute the intersections of the orbits with the fiducial plane, shown in gray, which contains the current LMC position and is orthogonal to its current velocity. All orbits cross this plane in nearby points, shown by crosses, but at different times, which in this case range between $-0.1$ and 0.1~Gyr, but the difference may be considerably larger. The coordinates of the crossing points in the fiducial plane form the spatial part of the transformed trajectory, while the time offset between the present-day position and the crossing point represents the remaining coordinate. In practice, orbits live in the 6d phase space and cross the fiducial 5d hyperplane, whereas the time offset is still the remaining sixth coordinate.
}  \label{fig:basis}
\end{figure}

This leads to the second step of coordinate transformation, illustrated in Figure~\ref{fig:basis}: each point in the 6d phase space gives rise to an orbit that crosses the fiducial hyperplane, and can be represented by the crossing point (coordinates in this 5d hyperplane) and the time of crossing. Unlike the first step, which was just a linear transformation effected by an orthogonal $6\times6$ matrix, the second transformation is manifestly nonlinear, and is also not quite symmetric. The conversion from physical to scaled coordinates $\boldsymbol u$ requires an entire orbit, not just a single point: we linearly transform the orbit into the rotated basis, determine the moment at which it crosses the fiducial hyperplane (i.e., when its first component in the rotated basis changes sign from negative to positive), and use the time offset between the actual point and the crossing point as the first coordinate $u_1$, and the remaining five coordinates of the crossing point in the rotated basis as $u_2..u_6$. The inverse transformation from scaled to physical space requires orbit integration from a point in the fiducial hyperplane specified by $u_2..u_6$ for a time $u_1$, and can be carried out only approximately (because the equations of motion ignore the residual accelerations not described by the assumed combined potential $\Phi$). Nevertheless, as long as $u_1$ is much shorter than dynamical time, this procedure remains sufficiently accurate to capture the primary source of nonlinearity (curvilinear trajectory in the potential $\Phi$).

\subsection{Iterative adjustment of orbital initial conditions}

To determine the direction in which the orbital initial conditions should be adjusted, at each iteration we perform a series of simulations with slightly different initial conditions, find the deviations of each orbit in the series from the target 6d phase-space point (present-day position and velocity of the LMC), and then use the Newton method to determine the next initial conditions that will hopefully bring the simulated orbit closer to the target. The initial and final deviations are considered in the nonlinearly transformed coordinates, as described in the previous section.

Let $\mathsf U$ be a $6\times (N+1)$ matrix of initial conditions in the basis associated with the initial phase-space point $\big\{ \boldsymbol x^{(0)}, \boldsymbol v^{(0)} \big\}$ of the master orbit, and $\mathsf W$ be a matrix of the same size with the final (present-day) phase-space coordinates in the basis associated with the true position and velocity of the LMC. As described above, the first element of each column vector in the nonlinearly transformed coordinates is the time offset between the point and the fiducial 5d hyperplane of the given basis, while the remaining five elements are the usual coordinates in this fiducial hyperplane. The first column $\boldsymbol u^{(0)}$ of the matrix $\mathsf U$ corresponds to the master orbit and is therefore zero by construction, but the corresponding first column $\boldsymbol w^{(0)}$ of the matrix $\mathsf W$ is not necessarily so -- in fact, making it zero is the goal of orbit finding. The usual assumption is that the mapping between the initial and the final conditions can be linearized in the vicinity of the master orbit, so that 
\begin{equation*}
\boldsymbol w^{(k)} \approx \boldsymbol w^{(0)} + \mathsf J \,\boldsymbol u^{(k)}
\end{equation*} 
for all companion orbits $k=1..N$, where $\mathsf J$ is the Jacobian of this mapping (which we also need to determine). This equality is only approximate for two reasons: first, a linearized form may not be valid for large displacements of companion orbits from the master one. Second, the final coordinates of orbits $\boldsymbol w^{(k)}$, including $\boldsymbol w^{(0)}$ itself, are measured with some uncertainty due to stochastic nature of $N$-body simulations. Therefore, the displacements of companion orbits should not be too small either, otherwise the Jacobian computed by subtracting final coordinates of nearby orbits will be dominated by noise. Because the master orbit is no better than others in terms of noisiness, it is advantageous to rewrite the above equation in terms of the deviation of all orbits from the target final coordinates (which are specified exactly -- of course, in reality the measurement of the LMC position and velocity has some observational uncertainty, but the goal of this procedure is to drive different simulations exactly to the same final point for a fair comparison). In the full matrix form, it reads
\begin{equation*}
\mathsf W \approx \mathsf J\, \mathsf U \,+\, \boldsymbol{\xi}\, \boldsymbol 1_{N+1}^\mathrm{T},
\end{equation*}
where the elements of the row vector $\boldsymbol 1_{N+1}^\mathrm{T}$ are ones, and the column vector $\boldsymbol{\xi}$ replaces $\boldsymbol w^{(0)}$ in the role of the final coordinates of the master orbit, and would coincide with it in the absence of noise. Both $\boldsymbol{\xi}$ and elements of the Jacobian matrix $\mathsf J$ are unknowns, and can be combined into a single matrix, while the matrix $\mathsf U$ is augmented with an extra row consisting of ones. Transposing these matrices, we have
\begin{equation*}
\bigg[ \boldsymbol 1_{N+1} \;\bigg|\; \mathsf U^T \bigg]  \cdot
\begin{bmatrix}
\boldsymbol \xi^\mathrm{T} \\
\mathsf J^\mathrm{T}
\end{bmatrix} \;\approx\;
\mathsf W^\mathrm{T}.
\end{equation*}
The first term on the left-hand side is a known design matrix of shape $(N+1)\times 7$, the second term is the matrix of unknowns of shape $7\times 6$, and the right-hand side is a matrix of shape $(N+1)\times 6$, measured with some errors. This kind of problem is naturally solved by a linear least-square fit with the same design matrix, but multiple columns on the right-hand side providing solutions for corresponding columns of the matrix of unknowns. We need $N+1$ to be at least 7, but a larger number of equations (overdetermined fit) will improve robustness of the inferred Jacobian and the final displacement vector of the master orbit. In practice, we use two companion orbits with opposite displacements for each coordinate, i.e., $N+1=13$.

 After obtaining the solution for $\mathsf J$ and $\boldsymbol\xi$, the next initial conditions are given by $\boldsymbol u^\mathrm{(next)} = -\mathsf J^{-1}\,\boldsymbol \xi$. However, if the Jacobian is ill-conditioned, the prescription for the next initial conditions may become quite unstable. This probably reflects a real physical degeneracy: one can make the initial orbital energy lower (case 1), or reduce its angular momentum while keeping energy fixed (case 2), and in the second case a smaller pericentre radius will create a stronger dynamical friction and cause the energy to drop faster, making the subsequent orbit very similar to case 1.

An alternative formulation deals with the vector of next initial conditions $\boldsymbol u^\mathrm{(next)}$ directly, writing the current initial conditions for all orbits as an inverse mapping from their final coordinates: 
\begin{equation*}
\boldsymbol u^{(k)} \approx \boldsymbol u^\mathrm{(next)} + \mathsf J^{-1} \boldsymbol w^{(k)}.
\end{equation*}
Now the matrix $\mathsf W$ is augmented with an extra row of ones, while the vector $\boldsymbol u^\mathrm{(next)}$ is combined with the inverse Jacobian into the matrix of unknowns in the least-square fit:
\begin{equation*}
\bigg[ \boldsymbol 1_{N+1} \;\bigg|\; \mathsf W^T \bigg]  \cdot
\begin{bmatrix}
\big( \boldsymbol u^\mathrm{(next)}\big)^\mathrm{T} \\
\big(\mathsf J^{-1}\big)^\mathrm{T}
\end{bmatrix} \;\approx\;
\mathsf U^\mathrm{T}.
\end{equation*}

The advantage of this approach is that the Jacobian is not used for inferring $\boldsymbol u^\mathrm{(next)}$, effectively become a set of nuisance parameters; however, now the design matrix $\mathsf W$ contains some noise, while the right-hand side is known perfectly, contrary to the assumptions of the least-square fit. Nevertheless, if the first procedure produces an unreasonable guess for the next initial conditions, we can use the more stable (even if generally less accurate) second formulation instead.

After the next initial conditions $\boldsymbol u^\mathrm{(next)}$ in the current basis are obtained, they are converted to physical coordinates by following the approximate equations of motion for the given time offset (the first element of the vector $\boldsymbol u$), and then a new master orbit and a new set of companion orbits are started from that point, with a new associated basis for converting physical to scaled coordinates. The iterative procedure typically needs to be repeated 3--5 times, with the early stages using lower-resolution simulations to save time (but sacrifice accuracy of computing the Jacobian).


\begin{thebibliography}{}

\bibitem[Astropy collaboration(2022)]{Astropy}
Astropy collaboration (Price-Whelan et al.), 2022, ApJ, 935, 167; ascl:1304.002
%definition of Galactocentric coordinates

\bibitem[Banik et al.(2022)]{Banik2022}
Banik I., Thies I., Truelove R., Candlish G., Famaey B., Pawlowski M., Ibata R., Kroupa P., 2022, MNRAS, 513, 129
%MW-M31 encounter in MOND

\bibitem[Battaglia et al.(2022)]{Battaglia2022}
Battaglia G., Taibi S., Thomas G., Fritz T., 2022, A\&A, 657, 54
%dSph PM from Gaia EDR3, LMC effect on satellite orbits

\bibitem[Besla et al.(2007)]{Besla2007}
Besla G., Kallivayalil N., Hernquist L., Robertson B., Cox T., van der Marel R., Alcock C., 2007, ApJ, 668, 949
%first approach scenario for LMC

\bibitem[Besla et al.(2010)]{Besla2010}
Besla G., Kallivayalil N., Hernquist L., van der Marel R., Cox T., Kere\v s D., 2010, ApJ, 721, L97
%Magellanic stream sim
 
\bibitem[B\'\i lek et al.(2018)]{Bilek2018}
B\'\i lek M., Thies I., Kroupa P., Famaey B., 2018, A\&A, 614, 59
%MW-M31 encounter in MOND

\bibitem[Boylan-Kolchin et al.(2011)]{BoylanKolchin2011}
Boylan-Kolchin M., Besla G., Hernquist L., 2011, MNRAS, 414, 1560
%LMC-SMC analogues in Millenium-2 sim

\bibitem[Busha et al.(2011)]{Busha2011}
Busha M., Wechsler R., Behroozi P., Gerke B., Klypin A., Primack J., 2011, ApJ, 743, 117
%LMC-MW analogues in SDSS and cosmo sims

\bibitem[Cautun et al.(2019)]{Cautun2019}
Cautun M., Deason A., Frenk C., McAlpine S., 2019, MNRAS, 483, 2185
%future of MW-LMC collision

\bibitem[Conroy et al.(2021)]{Conroy2021}
Conroy C., Naidu R., Garavito-Camargo N., Besla G., Zaritsky D., Bonaca A., Johnson B., 2021, Nature, 592, 534
%MW halo asymmetry due to LMC

\bibitem[Correa Magnus \& Vasiliev(2022)]{CorreaMagnus2022}
Correa Magnus L., Vasiliev E., 2022, MNRAS, 511, 2610
%LMC rewinding and MW mass estimate

\bibitem[Cunningham et al.(2020)]{Cunningham2020}
Cunningham E., Garavito-Camargo N., Deason A., et al., 2020, ApJ, 898, 4
%MW halo distortion due to LMC

\bibitem[Deason et al.(2015)]{Deason2015}
Deason A., Wetzel A., Garrison-Kimmel S., Belokurov V., 2015, MNRAS, 453, 3568
%LMC satellites from ELVIS sim

\bibitem[Deason et al.(2019)]{Deason2019}
Deason A., Belokurov V., Sanders J., 2019, MNRAS, 490, 3426
%MW stellar halo mass

\bibitem[Dehnen(2000)]{Dehnen2000}
Dehnen W., 2000, ApJ, 536, L9; ascl:1402.031
%gyrfalcON (N-body simulation code)

\bibitem[D'Onghia \& Lake(2008)]{DOnghia2008}
D'Onghia E., Lake G., 2008, ApJ, 686, L61
%group infall of satellites

\bibitem[Donaldson et al.(2022)]{Donaldson2022}
Donaldson K., Petersen M., Pe\~narrubia J., 2022, MNRAS, 513, 46
%LMC-MW simulation and the effect on local dark matter properties

\bibitem[Dooley et al.(2017)]{Dooley2017}
Dooley G., Peter A., Carlin J., Frebel A., Bechtol K., Willman B., 2017, MNRAS, 472, 1060
%missing satellites of LMC

\bibitem[Drimmel \& Poggio(2018)]{Drimmel2018}
Drimmel R., Poggio E., 2018, RNAAS, 2, 210
%solar velocity

\bibitem[D'Souza \& Bell(2022)]{DSouza2022}
D'Souza R., Bell E., 2022, MNRAS, 512, 739
%uncertainty in backward integration of orbits

\bibitem[Erkal et al.(2019)]{Erkal2019}
Erkal D., Belokurov V., Laporte C., et al., 2019, MNRAS, 487, 2685
%LMC mass from Orphan stream

\bibitem[Erkal \& Belokurov(2020)]{Erkal2020a}
Erkal D., Belokurov V., 2020, MNRAS, 495, 2554
%LMC satellites census

\bibitem[Erkal et al.(2020)]{Erkal2020b}
Erkal D., Belokurov V., Parkin D., 2020, MNRAS, 498, 5574
%LMC biases MW mass

\bibitem[Erkal et al.(2021)]{Erkal2021}
Erkal D., Deason A., Belokurov V., et al., 2021, MNRAS, 506, 2677
%MW halo kinematic asymmetries due to LMC

\bibitem[Fritz et al.(2018)]{Fritz2018}
Fritz T., Battaglia G., Pawlowski M., et al., 2018, A\&A, 619, 103
%dSph PM and orbits from Gaia DR2

\bibitem[Fritz et al.(2019)]{Fritz2019}
Fritz T., Carrera R., Battaglia G., Taibi S., 2019, A\&A, 623, 129
%LMC association for some ultrafaints

\bibitem[Gaia Collaboration(2018)]{Helmi2018}
\Gaia Collaboration (Helmi et al.), 2018, A\&A, 616, 12
%PM of Milky Way satellites

\bibitem[Gaia Collaboration(2021)]{Luri2021}
\Gaia Collaboration (Luri et al.), 2021, A\&A, 649, 7
%LMC in Gaia EDR3

\bibitem[Garavito-Camargo et al.(2019)]{GaravitoCamargo2019}
Garavito-Camargo N., Besla G., Laporte C.,  Johnston K., G\'omez F., Watkins L., 2019, ApJ, 884, 51
%LMC-MW simulation

\bibitem[Garavito-Camargo et al.(2021a)]{GaravitoCamargo2021a}
Garavito-Camargo N., Besla G., Laporte C., Price-Whelan A., Cunningham E., Johnston K., Weinberg M., G\'omez F., 2021, ApJ, 919, 109
%MW distortion due to LMC

\bibitem[Garavito-Camargo et al.(2021b)]{GaravitoCamargo2021b}
Garavito-Camargo N., Patel E., Besla G., Price-Whelan A., Gomez F., Laporte C., Johnston K., 2021, ApJ, 923, 140
%orbital pole clustering due to LMC

\bibitem[Garavito-Camargo et al.(2023)]{GaravitoCamargo2023}
Garavito-Camargo N., et al., 2023, submitted
%orbital pole clustering in cosmosim

\bibitem[G\'omez et al.(2015)]{Gomez2015}
G\'omez F., Besla G., Carpintero D., Villalobos \'A., O'Shea B., Bell E., 2015, ApJ, 802, 128
%MW and Sgr stream response to LMC

\bibitem[Guglielmo et al.(2014)]{Guglielmo2014}
Guglielmo M., Lewis G., Bland-Hawthorn J., 2014, MNRAS, 444, 1759
%LMC orbital history

\bibitem[Hansen \& Moore(2006)]{Hansen2006}
Hansen S., Moore B., 2006, New Astron., 11, 333
%velocity anisotropy in DM haloes

\bibitem[Hashimoto et al.(2003)]{Hashimoto2003}
Hashimoto Y., Funato Y., Makino J., 2003, ApJ, 582, 196
%distance-dependent Coulomb logarithm

\bibitem[Hasselquist et al.(2021)]{Hasselquist2021}
Hasselquist S., Hayes C., Lian J., et al., 2021, ApJ, 923, 172
%SFH of MW satellites

\bibitem[Hernquist \& Ostriker(1992)]{Hernquist1992}
Hernquist L., Ostriker J., 1992, ApJ, 386, 375
%basis set method

\bibitem[Jahn et al.(2019)]{Jahn2019}
Jahn E., Sales L., Wetzel A., Boylan-Kolchin M., Chan T., El-Badry K., Lazar A., Bullock J., 2019, MNRAS, 489, 5348
%LMC analogues and satellites in FIRE sims

\bibitem[Jeffreys(1952)]{Jeffreys1952}
Jeffreys H., 1952, Proc.R.Soc.London, series A, 214, 281
%Bakerain lecture: The origin of the Solar System

\bibitem[Jethwa et al.(2016)]{Jethwa2016}
Jethwa P., Erkal D., Belokurov V., 2016, MNRAS, 461, 2212
%LMC satellites and orbit rewinding

\bibitem[Kallivayalil et al.(2006)]{Kallivayalil2006a}
Kallivayalil N., van der Marel R., Alcock C., Axelrod T., Cook K., Drake A., Geha M., 2006, ApJ, 638, 772
%LMC PM from HST

\bibitem[Kallivayalil et al.(2013)]{Kallivayalil2013}
Kallivayalil N., van der Marel R., Besla G., Anderson J., Alcock C., 2013, ApJ, 764, 161
%updated LMC PM from HST and orbit reconstructions

\bibitem[Kallivayalil et al.(2018)]{Kallivayalil2018}
Kallivayalil N., Sales L., Zivick P., et al., 2018, ApJ, 867, 19
%LMC satellites in DR2

\bibitem[Kanehisa et al.(2023)]{Kanehisa2023}
Kanehisa K., Pawlowski M., M\"uller O., 2023, arXiv:2307.03218
%sat planes in Illustris

\bibitem[Koposov et al.(2009)]{Koposov2009}
Koposov S., Yoo J., Rix H.-W., Weinberg D., Macci\`o A., Miralda-Escud\'e J., 2009, ApJ, 696, 2179
%Milky Way satellite population incompleteness

\bibitem[Koposov et al.(2023)]{Koposov2023}
Koposov S., Erkal D., Li T.S., et al., 2023, MNRAS, 521, 4936
%Orphan stream + LMC

\bibitem[Kroupa et al.(2005)]{Kroupa2005}
Kroupa P., Theis C., Boily C., 2005, A\&A, 431, 517
%VPOS

\bibitem[Laporte et al.(2018a)]{Laporte2018a}
Laporte C., G\'omez F., Besla G., Johnston K., Garavito-Camargo N., 2018a, MNRAS, 473, 1218
%LMC impact on the MW disc

\bibitem[Laporte et al.(2018b)]{Laporte2018b}
Laporte C., Johnston K., G\'omez F., Garavito-Camargo N., Besla G., 2018b, MNRAS, 481, 286
%Sgr and LMC impact on the MW disc

\bibitem[Lemasle et al.(2022)]{Lemasle2022}
Lemasle B., Lala H., Kovtyukh V., et al., 2022, A\&A, 668, 40
%MW disc warp in cepheids

\bibitem[Li \& Helmi(2008)]{Li2008}
Li Y.-S., Helmi A., 2008, MNRAS, 385, 1365
%group accretion of satellites

\bibitem[Lilleengen et al.(2023)]{Lilleengen2023}
Lilleengen S., Petersen M., Erkal D., et al., 2023, MNRAS, 518, 774
%simulation of deforming LMC-MW

\bibitem[Lokas et al.(2014)]{Lokas2014}
Lokas E., Athanassoula E., Debattista V., Valluri M., del Pino A., Semczuk M., Gajda G., Kowalczyk K., 2014, MNRAS, 445, 1339
%tidally induced bar

\bibitem[Lowing et al.(2011)]{Lowing2011}
Lowing B., Jenkins A., Eke V., Frenk C., 2011, MNRAS, 416, 2697
%basis-set halo expansion

\bibitem[Lynden-Bell(1976)]{LyndenBell1976}
Lynden-Bell D., 1976, MNRAS, 174, 695
%Magellanic plane and satellites: Draco, UMi, Sculptor; Pal 1, Pal 13

\bibitem[Lynden-Bell(1982a)]{LyndenBell1982a}
Lynden-Bell D., 1982a, The Observatory, 102, 7
%Magellanic groupe: UMi, Draco, Carina, Sculptor

\bibitem[Lynden-Bell(1982b)]{LyndenBell1982b}
Lynden-Bell D., 1982b, The Observatory, 102, 202
%Magellanic plane: Fornax, Sculptor, Leo I, Leo II

\bibitem[Majewski(1994)]{Majewski1994}
Majewski S., 1994, ApJ, 431, L17
%Sextans and Phoenix dSph as part of the FLS group

\bibitem[Makarov et al.(2023)]{Makarov2023}
Makarov D., Khoperskov S., Makarov D., Makarova L., Libeskind N., Salomon J.-B., 2023, MNRAS, 521, 3540
%MW satellites velocity apex

\bibitem[Massana et al.(2022)]{Massana2022}
Massana P., Ruiz-Lara T., No\"el N., et al., 2022, MNRAS, 513, L40
%SFH of LMC & SMC

\bibitem[Mathewson et al.(1974)]{Mathewson1974}
Mathewson D., Cleary M., Murray J., 1974, ApJ, 190, 291
%Magellanic stream

\bibitem[Mazzi et al.(2021)]{Mazzi2021}
Mazzi A., Girardi L., Zaggia S., et al., 2021, MNRAS, 508, 245
%SFH of LMC

\bibitem[McConnachie(2012)]{McConnachie2012}
McConnachie A., 2012, ApJ, 144, 4
%dSph database

\bibitem[Metz et al.(2008)]{Metz2008}
Metz M., Kroupa P., Libeskind N., 2008, ApJ, 680, 287
%kinematically coherent VPOS

\bibitem[Metz et al.(2009)]{Metz2009}
Metz M., Kroupa P., Theis C., Hensler G., Jerjen H., 2009, ApJ, 697, 269
%MW satellite disc unlikely to be result of group infall

\bibitem[Milgrom(1983)]{Milgrom1983}
Milgrom M., 1983, ApJ, 270, 365
%MOND

\bibitem[Nichols et al.(2011)]{Nichols2011}
Nichols M., Colless J., Colless M., Bland-Hawthorn J., 2011, ApJ, 742, 110
%Magellanic group accretion

\bibitem[Pace et al.(2022)]{Pace2022}
Pace A., Erkal D., Li T., 2022, ApJ, 940, 136
%orbits of MW and LMC satellites

\bibitem[Pardy et al.(2020)]{Pardy2020}
Pardy S., D'Onghia E., Navarro J., et al., 2020, MNRAS, 492, 1543
%suggest that Carina and Fornax are LMC sats

\bibitem[Patel et al.(2020)]{Patel2020}
Patel E., Kallivayalil N., Garavito-Camargo N., et al., 2020, ApJ, 893, 121
%orbits and classification of LMC satellites

\bibitem[Pawlowski et al.(2011)]{Pawlowski2011}
Pawlowski M., Kroupa P., de Boer K., 2011, A\&A, 532, 118
%tidal dwarf origin of satellite planes

\bibitem[Pawlowski et al.(2012)]{Pawlowski2012}
Pawlowski M., Pflamm-Altenburg J., Kroupa P., 2012, MNRAS, 423, 1109
%VPOS

\bibitem[Pawlowski \& Kroupa(2013)]{Pawlowski2013}
Pawlowski M., Kroupa P., 2013, MNRAS, 435, 2116
%VPOS kinematics

\bibitem[Pawlowski \& Kroupa(2020)]{Pawlowski2020}
Pawlowski M., Kroupa P., 2020, MNRAS, 491, 3042
%VPOS after Gaia DR2

\bibitem[Pawlowski(2021)]{Pawlowski2021}
Pawlowski M., 2021, Galaxies, 9, 66
%satellite planes review

\bibitem[Pawlowski et al.(2022)]{Pawlowski2022}
Pawlowski M., Oria P.-A., Taibi S., Famaey B., Ibata R., 2021, ApJ, 932, 70
%no effect of LMC on VPOS

\bibitem[Petersen \& Pe\~narrubia(2020)]{Petersen2020}
Petersen M., Pe\~narrubia J., 2020, MNRAS, 494, L11
%predicted signatures of LMC reflex motion

\bibitem[Petersen \& Pe\~narrubia(2021)]{Petersen2021}
Petersen M., Pe\~narrubia J., 2021, Nature Astronomy, 5, 251
%detection of LMC-induced MW reflex motion in halo kinematics

\bibitem[Pietrzy\'nski et al.(2019)]{Pietrzynski2019}
Pietrzy\'nski G., Graczyk D., Gallenne A., et al., 2019, Nature, 567, 200
%LMC distance

\bibitem[Price-Whelan(2017)]{PriceWhelan2017}
Price-Whelan A., 2017, JOSS, 2, 388; ascl:1707.006
%Gala

\bibitem[Romero-G\'omez et al.(2019)]{RomeroGomez2019}
Romero-G\'omez M., Mateu C., Aguilar L., Figueras F., Castro-Ginard A., 2019, A\&A, 627, 150
%MW disc warp in Gaia DR2

\bibitem[Rozier et al.(2022)]{Rozier2022}
Rozier S., Famaey B., Siebert A., Monari G., Pichon C., Ibata R., 2022, ApJ, 933, 113
%linear response models of LMC-MW interaction

\bibitem[Sales et al.(2011)]{Sales2011}
Sales L., Navarro J., Cooper A., White S., Frenk C., Helmi A., 2011, MNRAS, 418, 648
%LMC analogues and satellites in Aquarius sim

\bibitem[Sales et al.(2017)]{Sales2017}
Sales L., Navarro J., Kallivayalil N., Frenk C., 2017, MNRAS, 465, 1879
%LMC satellites

\bibitem[Samuel et al.(2021)]{Samuel2021}
Samuel J., Wetzel A., Chapman S., Tollerud E., Hopkins P., Boylan-Kolchin M., Bailin J., Faucher-Gigu\`ere C.-A., 2021, MNRAS, 504, 1379
%satellite planes in FIRE sim

\bibitem[Sanders et al.(2020)]{Sanders2020}
Sanders J., Lilley E., Vasiliev E., Evans N.W., Erkal D., 2020, MNRAS, 499, 4973
%halo potential expansion

\bibitem[Santos-Santos et al.(2021)]{SantosSantos2021}
Santos-Santos I., Fattahi A., Sales L., Navarro J., 2021, MNRAS, 504, 4551
%LMC analogues and satellites in Apostle sim

\bibitem[Smith et al.(2016)]{Smith2016}
Smith R., Duc P.-A., Bournaud F., Yi S., 2016, ApJ, 818, 11
%formation of satellite planes in mergers

\bibitem[Tepper-Garc\'\i a et al.(2019)]{TepperGarcia2019}
Tepper-Garci\'\i a T., Bland-Hawthorn J., Pawlowski M., Fritz T., 2019, MNRAS, 488, 918
%hydro sims of Magellanic stream

\bibitem[Tremaine(1976)]{Tremaine1976}
Tremaine S., 1976, ApJ, 203, 72
%LMC orbit with dynamical friction

\bibitem[van der Marel et al.(2002)]{vanderMarel2002}
van der Marel R., Alves D., Hardy E., Suntzeff N., 2002, AJ, 124, 2639
%LMC Vlos

\bibitem[van der Marel \& Kallivayalil(2014)]{vanderMarel2014}
van der Marel R., Kallivayalil N., 2014, ApJ, 781, 121
%LMC internal kinematics and mass estimate

\bibitem[Vasiliev(2018)]{Vasiliev2018}
Vasiliev E., 2018, MNRAS, 481, L100
%LMC dynamical model and mass estimate

\bibitem[Vasiliev(2019)]{Vasiliev2019}
Vasiliev E., 2019, MNRAS, 482, 1525; ascl:1805.008
%AGAMA

\bibitem[Vasiliev \& Belokurov(2020)]{Vasiliev2020}
Vasiliev E., Belokurov V., 2020, MNRAS, 497, 4162
%Sgr remnant, orbit IC fitting

\bibitem[Vasiliev et al.(2021)]{Vasiliev2021}
Vasiliev E., Belokurov V., Erkal D., 2021, MNRAS, 501, 2279
%Sgr stream + LMC

\bibitem[Vasiliev et al.(2022)]{Vasiliev2022}
Vasiliev E., Belokurov V., Evans N.~W., 2022, ApJ, 926, 203
%radialization of massive satellite orbits, implications for LMC rewinding

\bibitem[Vasiliev(2023)]{Vasiliev2023}
Vasiliev E., 2023, Galaxies, 11, 59 [V23]
%review of LMC-MW interaction

\bibitem[Wang et al.(2020)]{Wang2020}
Wang W., Han J., Cautun M., Li Z., Ishigaki M., 2020, SCPMA, 63, 109801
%review of MW mass profile estimates

\bibitem[Yozin \& Bekki(2015)]{Yozin2015}
Yozin C., Bekki K., 2015, MNRAS, 453, 2302
%UFD accreted with LMC

\end{thebibliography}
\end{document}